\documentclass{bmcart}
\usepackage[center]{subfigure}
\usepackage{bigints}
\usepackage{cancel}
\usepackage{amsmath}
\usepackage{mathbbol}
\usepackage[center]{subfigure}
\usepackage{tikz}
\usepackage{standalone}
\usepackage{hyperref}

\usepackage{mathtools}

\usepackage{amsthm,amsmath}
\usepackage[center]{subfigure}

\usepackage[utf8]{inputenc} %unicode support
\startlocaldefs
\endlocaldefs

\begin{document}

%%% Start of article front matter
\begin{frontmatter}

\begin{fmbox}
\dochead{Research}


\title{A Framework for Reconstructing COVID-19 Transmission Network to Inform Betweenness Centrality-Based Control Measures}

%%%%%%%%%%%%%%%%%%%%%%%%%%%%%%%%%%%%%%%%%%%%%%
%%                                          %%
%% Enter the authors here                   %%
%%                                          %%
%% Specify information, if available,       %%
%% in the form:                             %%
%%   <key>={<id1>,<id2>}                    %%
%%   <key>=                                 %%
%% Comment or delete the keys which are     %%
%% not used. Repeat \author command as much %%
%% as required.                             %%
%%                                          %%
%%%%%%%%%%%%%%%%%%%%%%%%%%%%%%%%%%%%%%%%%%%%%%

\author[
  addressref={aff1,aff2},                   % id's of addresses, e.g. {aff1,aff2}
  corref={aff1},                       % id of corresponding address, if any
% noteref={n1},                        % id's of article notes, if any
  email={sn62@aub.edu.lb}   % email address
]{\inits{S.N}\fnm{Sara} \snm{Najem}}
\author[
  addressref={aff1,aff3}
]{\inits{S.M}\fnm{Stefano} \snm{Monni}}
\author[
  addressref={aff2}
]{\inits{R.H}\fnm{Rola} \snm{Hatoum}}
\author[
  addressref={aff4}
]{\inits{H.S}\fnm{Hawraa } \snm{Sweidan}}
\author[
  addressref={aff5}
]{\inits{G.F}\fnm{Ghaleb} \snm{Faour}}
\author[
  addressref={aff5}
]{\inits{C.A}\fnm{Chadi} \snm{Abdallah}}
\author[
  addressref={aff4}
]{\inits{N.G}\fnm{Nada} \snm{Ghosn}}
\author[
  addressref={aff6}
]{\inits{H.H}\fnm{Hamad} \snm{Hassan}}
\author[
  addressref={aff1,aff2},                   % id's of addresses, e.g. {aff1,aff2}
  corref={aff1},                       % id of corresponding address, if any
% noteref={n1},                        % id's of article notes, if any
  email={jtt00@aub.edu.lb}   % email address
]{\inits{J.T}\fnm{Jihad} \snm{Touma}}

%%%%%%%%%%%%%%%%%%%%%%%%%%%%%%%%%%%%%%%%%%%%%%
%%                                          %%
%% Enter the authors' addresses here        %%
%%                                          %%
%% Repeat \address commands as much as      %%
%% required.                                %%
%%                                          %%
%%%%%%%%%%%%%%%%%%%%%%%%%%%%%%%%%%%%%%%%%%%%%%

\address[id=aff1]{%                           % unique id
  \orgdiv{Department of Physics},             % department, if any
  \orgname{ American University of Beirut},          % university, etc
  \city{Beirut},                              % city
  \cny{Lebanon}                                    % country
}
\address[id=aff2]{%
  \orgdiv{Center for Advanced Mathematical Sciences},
  \orgname{American University of Beirut},
  %\street{},
  %\postcode{}
  \city{Beirut},
  \cny{Lebanon}
}

\address[id=aff3]{%
  \orgdiv{Department of Mathematics},
  \orgname{American University of Beirut},
  %\street{},
  %\postcode{}
  \city{Beirut},
  \cny{Lebanon}
}
\address[id=aff4]{%
  \orgdiv{Epidemiological Surveillance Program},
  \orgname{Ministry of Public Health},
  %\street{},
  %\postcode{}
  \city{Beirut},
  \cny{Lebanon}
}
\address[id=aff5]{%
  \orgdiv{National Center for Remote Sensing},
  \orgname{National Council for Scientific Research (CNRS)},
  %\street{},
  %\postcode{}
  \city{Beirut},
  \cny{Lebanon}
}

\address[id=aff6]{%
  \orgdiv{Faculty of Public Health},
  \orgname{Lebanese University},
  %\street{},
  %\postcode{}
  \city{Beirut},
  \cny{Lebanon}
}
%\author{Stefano Monni$^{2,3}$ }
%\author{Rola Hatoum$^{2}$}
%\author{Hawraa Sweidan$^{4}$}
%\author{Ghaleb Faour$^{5}$}
%% \author[5]{Chadi Abdallah}
%\author{Nada Ghosn$^{4}$}
%\author{Hamad Hassan$^{6}$}
%\author{Jihad Touma$^{1,2,*}$ }
%\affiliation{$^1$ Department of Physics, American University of Beirut, Beirut, Lebanon}
%\affiliation{$^2$ Center for Advanced Mathematical Sciences, American University of Beirut, Beirut, Lebanon}
%\affiliation{$^3$Department of Mathematics, American University of Beirut, Beirut,  Lebanon} %1107 2020,
%\affiliation{$^4$Epidemiological Surveillance Program, Ministry of Public Health, Beirut, Lebanon} %1107 2020,
%\affiliation{$^5$National Center for Remote Sensing, National Council for Scientific Research (CNRS), Riad al Soloh, Beirut, Lebanon} %1107 2020,
%\affiliation{$^6$Faculty of Public Health, Lebanese University,  Lebanon}
%\affiliation{$^*$corresponding authors: Sara Najem (sn62@aub.edu.lb), Jihad Touma (jt00@aub.edu.lb)}

%%%%%%%%%%%%%%%%%%%%%%%%%%%%%%%%%%%%%%%%%%%%%%
%%                                          %%
%% Enter short notes here                   %%
%%                                          %%
%% Short notes will be after addresses      %%
%% on first page.                           %%
%%                                          %%
%%%%%%%%%%%%%%%%%%%%%%%%%%%%%%%%%%%%%%%%%%%%%%

%%\note{Sample of title note}     % note to the article
%\note[id=n1]{Equal contributor} % note, connected to author

\end{fmbox}% comment this for two column layout

%%%%%%%%%%%%%%%%%%%%%%%%%%%%%%%%%%%%%%%%%%%%%%%
%%                                           %%
%% The Abstract begins here                  %%
%%                                           %%
%% Please refer to the Instructions for      %%
%% authors on https://www.biomedcentral.com/ %%
%% and include the section headings          %%
%% accordingly for your article type.        %%
%%                                           %%
%%%%%%%%%%%%%%%%%%%%%%%%%%%%%%%%%%%%%%%%%%%%%%%

\begin{abstractbox}

\begin{abstract} % abstract
In this paper, we propose a general framework for optimal control measures, which follows the evolution of COVID-19 infection counts collected by Surveillance Units on a country level. 
 We employ an autoregressive model that allows to decompose the mean number of infections into three components that describe: intra-locality infections, inter-locality infections, and infections from other sources such as travelers arriving to a country from abroad. We identify the inter-locality term as a time-evolving network and when it drives the dynamics of the disease we focus on its properties. Tools from network analysis are then employed to get insight into its topology. Building on this, and particularly on the centrality of the nodes of the identified network, a strategy for intervention and disease control is devised. 
 \end{abstract}

%%%%%%%%%%%%%%%%%%%%%%%%%%%%%%%%%%%%%%%%%%%%%%
%%                                          %%
%% The keywords begin here                  %%
%%                                          %%
%% Put each keyword in separate \kwd{}.     %%
%%                                          %%
%%%%%%%%%%%%%%%%%%%%%%%%%%%%%%%%%%%%%%%%%%%%%%


% MSC classifications codes, if any
%\begin{keyword}[class=AMS]
%\kwd[Primary ]{}
%\kwd{}
%\kwd[; secondary ]{}
%\end{keyword}

\end{abstractbox}
%
%\end{fmbox}% uncomment this for two column layout

\end{frontmatter}

%%%%%%%%%%%%%%%%%%%%%%%%%%%%%%%%%%%%%%%%%%%%%%%%
%%                                            %%
%% The Main Body begins here                  %%
%%                                            %%
%% Please refer to the instructions for       %%
%% authors on:                                %%
%% https://www.biomedcentral.com/getpublished %%
%% and include the section headings           %%
%% accordingly for your article type.         %%
%%                                            %%
%% See the Results and Discussion section     %%
%% for details on how to create sub-sections  %%
%%                                            %%
%% use \cite{...} to cite references          %%
%%  \cite{koon} and                           %%
%%  \cite{oreg,khar,zvai,xjon,schn,pond}      %%
%%                                            %%
%%%%%%%%%%%%%%%%%%%%%%%%%%%%%%%%%%%%%%%%%%%%%%%%

%%%%%%%%%%%%%%%%%%%%%%%%% start of article main body
% <put your article body there>

%%%%%%%%%%%%%%%%
%% Background %%
%%
\section{Introduction}
%\subsection{}
\noindent
Many studies have appeared in the literature to address a wealth of questions about
the COVID-19 pandemic, covering a range of disciplines  from  medicine, genetics, pharmacology to social and economic sciences \cite{mohamadou2020review,guan2020modeling}.     Some of these studies have a regional focus, investigating the disease in countries, or  regions of a country, while others have considered the pandemic in  larger geographical contexts \cite{celani2021endemic,ssentongo2021pan,dickson2020assessing}. 

Modeling  the evolution of the disease  has been one of the main quantitative approaches used, especially with the goal of forecasting   numbers of affected cases. It has also been at the heart of subsequent investigations on practical aspects:  from  management of resources (assessment of the preparedness for disease containment and  readiness of the healthcare system) to possible intervention measures (vaccination  and testing strategies \cite{gozzi2021importance}, government control measures)  and their consequences \cite{perra2021non}.
The different  types of modeling  that   have been applied to investigate the dynamics of  COVID-19 infections have  a long history in epidemic modeling.  Compartmental models ({\it e.g.}, SIR, SIER) are a well known class of models. They study the interplay between susceptible, infected and recovered individuals  within  communities, with different degrees of spatial refinement. For instance, in the so called networked compartmental models,  interactions between communities are encoded in a network \cite{brockmann2013hidden},  often to identify the  spatial and temporal origin of the disease \cite{schlosser2021finding}. In statistics, spatial/and/or temporal point processes  are often employed  to study the  dynamics of the disease.
% number of
% infected cases in a specific geographical region. 
Some models allow for the number of infections to be triggered by those at previous times, others can  incorporate, as covariates, additional available information  such as demographics,  human mobility, and policy decisions. Some recent work on COVID-19 follows this direction  \cite{zhu2021high, chiang2021hawkes, giudici2021network}.  Epidemic models can also be recast in a standard regression framework, where the time series of infection counts are fitted by specifying a distribution for the counts and the associated conditional mean function \cite{hhh, meyer2014spatio}.  In most of these studies, the goal is to predict and capture the spatio-temporal patterns of disease spread. Recently, this approach was implemented on  COVID-19 data \cite{celani2021endemic, ssentongo2021pan, dickson2020assessing}.
{In this paper, we  propose a framework for optimal control measures. The first component of the framework hinges on a statistical  regression analysis of infection counts over time and aggregated over localities in a country, with a mean function that takes account of the spatial proximity of these localities. The fitted model is then used to reconstruct a weighted network, which constitutes the second component of our framework.
The salient point of our method of network recovery is that  smoothness conditions on the temporal data are not required \cite{shandilya2011inferring}, and neither is the near steady-state dynamics  that is instead necessary for  the perturbation/response approaches to work \cite{prabakaran2014paradoxical,yu2010estimating,yu2010inferring}. Our reconstruction is similar to that  in \cite{wan2014inferring}, while it differs from  procedures that rely on the deterministic evolution of the disease  \cite{pajevic2009efficient,braunstein2019network}. The third component of our framework is the study of  the changes in the topology of the underlying recovered network and the computation of centrality measures (specifically the nodes' betweenness centrality), from which the recommendation of optimal control measures ensues. We apply the method in the case of Lebanon.

%is an analysis of the COVID-19 data using  regression modeling.  %Network information encoding the distances among localities is used in the mean function, similarly to  \cite{meyer2014power}, to model  the inter-locality contribution to the mean number of infection. %  We interpret the corresponding estimates as a network which is then  studied using network tools as a function of time.
% The combination of these approaches  is  then brought to bear  to guide optimal control strategies.
% we make an observation about the network encoding the interactions between localities and study is using network analyses tools. Network information encoding the distances among localities is used in the model function.  This information  has previously been used in  time series  models for infectious disease spread \cite{meyer2014power}.  The novelty of our approach is to c
 }
\section{Model Definition}
% The first cases of  COVID-19 in Lebanon were recorded in February 2020. 

Our starting  point of the analysis is a statistical model that captures  the spatio-temporal dynamics of the infections, under the statistical framework discussed in \cite{hhh}. 
Namely, we consider the number $Y_{it}$ of infections  recorded  in a given locality $i$ in a given day $t$,  as independent, conditionally on the counts at previous times, random variables  distributed according to a  negative binomial distribution having a  mean function decomposed into three terms as follows:
\begin{equation}\label{autoreg}
E(Y_{i,t}|Y_{j,t-1}, e_i)=\mu_{it}=  \lambda_{it}Y_{i,t-1} +  \phi_{it} \sum_{j \neq i}  \omega_{ij}Y_{j,t-1} +\nu_{it}e_i.
\end{equation}
The first two terms constitute the auto-regressive part of the model: one being the contribution to the mean infection $\mu_{it}$ in locality $i$ at time $t$,  due to the infections  within $i$ at the previous day,  the other being the contribution to $\mu_{it}$ due to positive cases from other localities $j$ also at the previous day. The final term accounts for all other contributions not captured by the first two, such as infected people who entered the country under study  from abroad. %through the airport. 
% It is worth mentioning here that the data do not include information on contact tracing, accordingly in our approach we do not attempt to recover the contact network. We are rather trying to infer a coarser transmission network that involves effects between localities rather than between individuals.
For simplicity we will refer to the last term $\nu_{it}e_i$ as the component due to travel and assume that it is proportional to the size of the population  $e_i$ of the locality. 
The log-transforms of  non-negative coefficients  $\lambda_{it}$ and    $\phi_{it}$, which quantify  the contribution of the past observations to future counts, and the log-transform of  the  parameter $\nu_{it}$   are each modeled as  a  linear function  of time,  with a locality-specific slope to allow more flexibility across localities. Intercepts and slopes are estimated from the data. Finally, we model $\omega_{ji}$ as a power function of the geographical distance $d_{ij}$ of the localities: $\omega_{ji}  \propto  d_{ij}^{-f}$. This is assumed because previous studies have shown that  mobility flows are governed by power-law functions of inter-localities distances \cite{meyer2014power,zipf1946p,simini2012universal,barthelemy2011spatial}.
\section{Data Description and Fitting of the Regression Model}

As an application of the framework, the model (\ref{autoreg}) was applied to the COVID-19 data collected in Lebanon. On daily basis, the laboratories from the public and private sectors report the confirmed cases to the Epidemiological Surveillance Program of Lebanon's Ministry of Public Health (ESUMOH). Later, the cases are investigated in order to get additional demographic information and health condition. The data are then archived in a national platform.
% On a weekly basis, a dataset including the date of reporting and the locality of residence is provided for analysis.
% Since then, new cases have been recorded every day. In this work, we analyse the number of cases from that first record until present. 
Specifically, the data we have considered consist of counts of COVID-19  recorded daily  in   each of the 1544  localities of  Lebanon from February 21, 2020 to January 20, 2022.  Such localities correspond to Lebanon's smallest statistical units called ``circonscriptions fonci\`eres" or cadastral villages following the Central Administration for Statistics (CAS) nomenclature \cite{verdeil2007atlas}. 
Recommendations on possible interventions and updates on the disease evolution were sought for by the Ministry of Public Health at 20 days intervals. Model  (\ref{autoreg})  was fitted using the R  package {\sf surveillance} \cite{meyer2014spatio} over the intervals $[0,t]$, $t=20 \cdot n$ for $n=1, \cdots, 35$. This allows us to follow the evolution in time of the model parameters until day 700, the last observation point. 

\begin{figure}[!htp]
\includegraphics[width=0.5\textwidth]{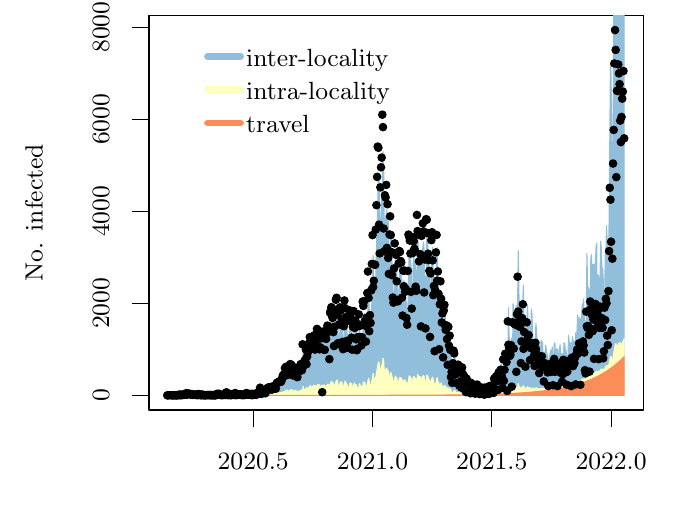}
\caption{Data and fit under the model  of Equation (\ref{autoreg}) with  negative binomial distributed counts.  The three colors show the  decomposition of the fitted aggregate counts into travel, intra-locality, and inter-locality contributions to infections amounting to $3\%$, $10\%$, and $87\%$ respectively. }
\label{fig:modelall}       % Give a unique label
\end{figure}

Figure  \ref{fig:modelall} displays   the aggregated counts over all localities, $\sum_{i} Y_{i, t}$, and the fitted values over the complete time period of our study broken down into  the three components of the mean function. The fit appears quite adequate. It is in fact a better fit to the data than the model with counts assumed to be Poisson-distributed, which is an indication of overdispersion in the data. A further comparison of these two models in terms of AIC value and prediction errors is provided in Table  \ref{table:poivnbin}.

\begin{table}[!htp] 
\begin{tabular}{@{}|lcc|@{}}
\hline
 & ses & AIC \\ \hline
Negative Binomial   & 2.88 & 1164119 \\ 
Poisson &  3.65 & 1333453 \\ 
\hline
\end{tabular}
\caption{Comparison of the two models with the same mean function given in equation (\ref{autoreg})  with distribution of the counts being negative binomial and Poisson.
AIC is Akaike's Information criterion. SES is the mean squared error: the mean squared  difference over the localities of the  observed and predicted  counts at the final time point of the study}\label{table:poivnbin}
%obtained from the model trained on  data at previous times over the localities).  }
\end{table}
We further notice that the model of equation (\ref{autoreg}) considers  a time-lag of one day; that is, the future counts depend on the counts recorded on the previous day. Changes in the time-lag from one to a few days did not result in any
noticeable difference. 
Our analysis provides evidence that the inter-locality infection  drives the overall  transmission of the disease \cite{meyer2014power}. Then, for this reason we shift our focus to the network that governs the interaction between localities and observe that it is not purely a static spatially-dependent network but rather dynamic and time-evolving: in fact the product of time-dependent coefficients with the spatial proximity matrix.
Figure  \ref{fig:modelall}  indicates that the inter-locality term  has the most important contribution  to the increase in the mean number of infections compared to the intra-locality and travel terms. This suggests that the inter-locality transmissions should be the main focus of  analysis, and  what one learns from  their study  would be useful for disease control. 

\begin{figure}[!htp]
\includegraphics[width=0.5\textwidth]{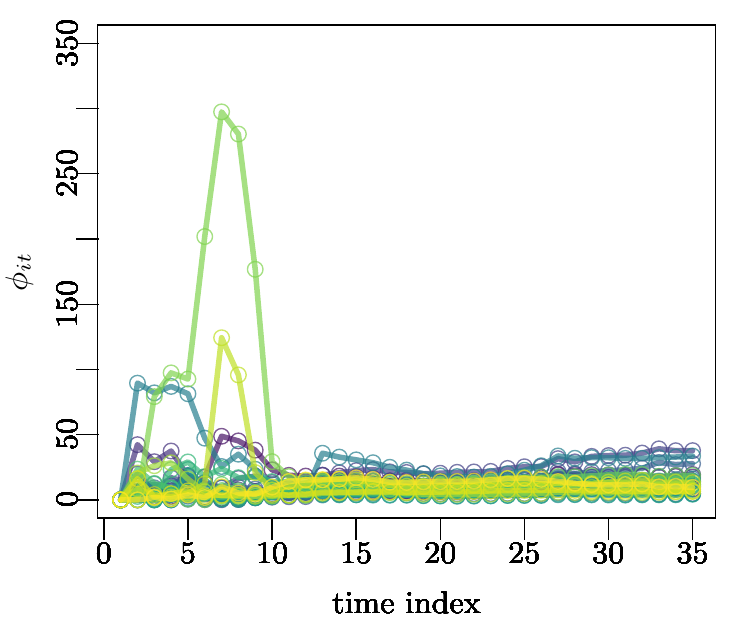}
\caption{Time evolution of the inter-locality parameter $\phi_{it}$ is shown for the regions with the highest centrality (see section \ref{sec:net_char}).}
\label{fig:phi} 
\end{figure}

The parameter estimates of model  (\ref{autoreg}) for all 1554 localities and their errors can not be displayed in an uncluttered fashion but are available from the corresponding authors. A sample of the evolution of the inter-locality term $\phi_{it}$ is shown in Figure \ref{fig:phi}.

\section{Network Identification and  Characterization}\label{sec:net_char}

\begin{figure}[!htp]
\includegraphics[width=0.5\textwidth]{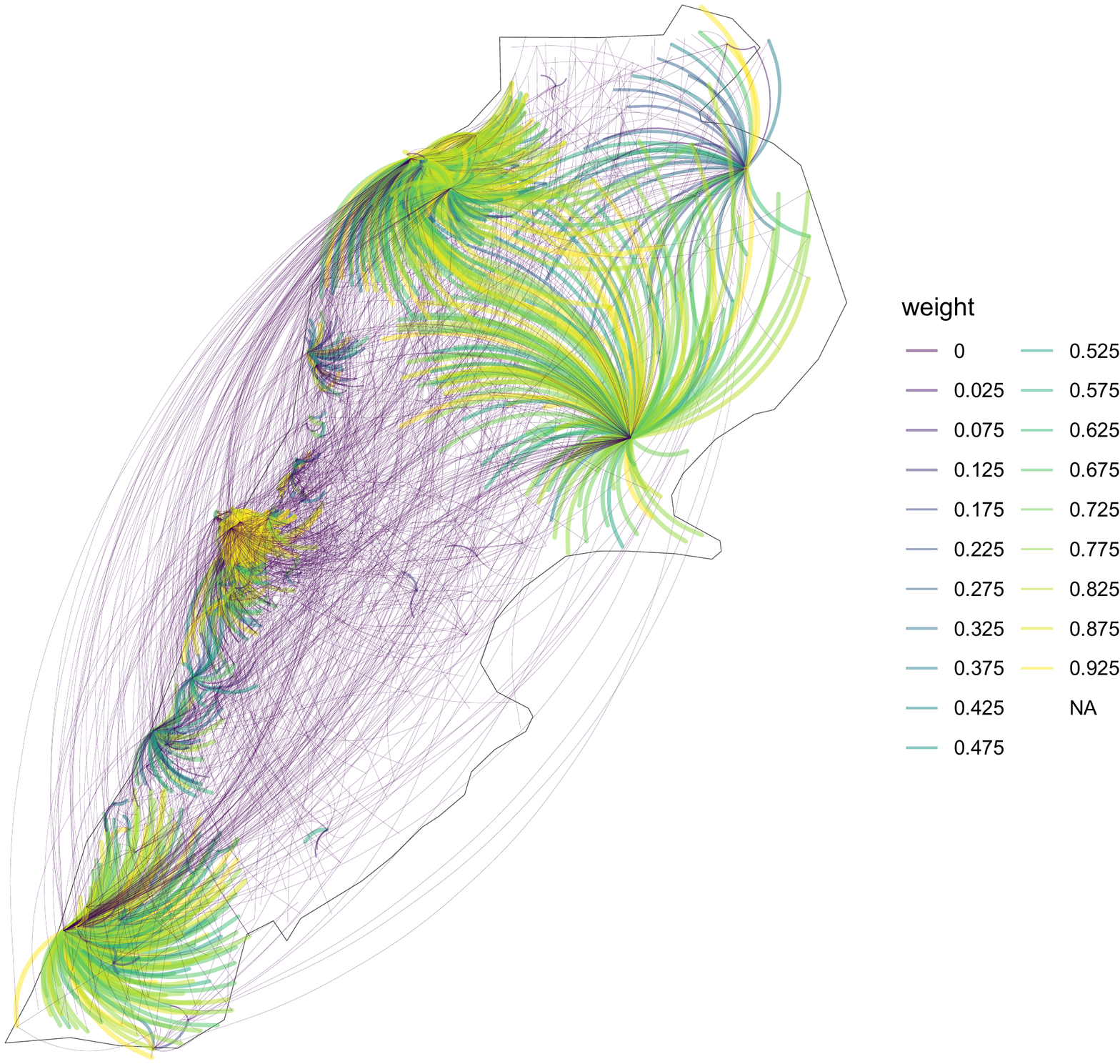}
\caption{The network $A^{(15)}(t=300)$  is overlaid on the map of Lebanon to illustrate its complexity. It is evaluated at the 300-th day of the pandemic using all 15 time interval of our collected data, that is the time span $[0,20\cdot 15]$. }
\label{fig:amtrix}       % Give a unique label
\end{figure}

% {\bf Models such as  \label{autoreg}, with a term that takes  account of  the mobility information,  has been previously used, e.g.  \cite{meyer2014power},  for infectious disease spread.  What we  now present here is }
\noindent
As has emerged from the statistical model, the inter-locality dynamics plays the major role
in  determining the infection  numbers. While this may seem to contradict previous studies, for example \cite{karumanagoundar2021secondary}, on the secondary attack rate being mostly driven by household interactions, it is worth noting that infections outside households are hard to pin down, and thus this may be a limitation in such studies. We wish to focus now on the inter-locality term and study it from a different perspective. To do so, observe that the second  term in the  mean equation (\ref{autoreg}) can be re-written as follows: $
\sum_{j \neq i}A_{ij}(t) Y_{j, t-1}$, where $A_{ij}(t)=\phi_{it}\cdot  \omega_{ij}$. It can be interpreted as the contribution to the cases at time $t$ in locality $i$ from cases from locality $j$ at the previous day.
We can suggestively think of   $A=(A_{ij})$ as defining the weights of a network  between localities:  the transport network $\omega$ describes  the traffic flow between localities, and thus predates the disease, while $\phi_{it}$ is  the number of transported cases from
 $i$ into neighboring localities.  %The network $A$ is estimated on some training data from $[0,t]$ and can be evaluated at any time $t'$. 
 $A(t)$ explicitly depends on $t$ since the coefficients of $\log \phi_{it}$, which are linear functions of $t$, and the power $f$  in the definition of $\omega$ are  estimated   over each interval $[0,t]$.   An example is provided in Figure \ref{fig:amtrix} which is  a graphical representation of  $A^{(15)}(t=300)$. The superscript $15$ in the notation of $A$ indicates that the latter was estimated on the counts data of $15$  contiguous 20-day time intervals, that is the 300-day time span $[0,20\cdot 15]$ from February 21, 2020 to December 16th, 2020. 
%  , that is, the evaluation of $A$ is at  300 days into the pandemic. For this figure,  the training data were the counts on the first 300 days. 
 %The edges are color-coded according to their weights $A_{ij}$,  with lighter  shades  indicating  larger values and the color white indicating absence of edges. 
 This complex network $A$ drives the cross-localities dynamics.  We will now suggest employing some useful summary metrics for $A(t)$ and its time evolution in order to understand its properties, and accordingly prescribe adequate control measures. 

\begin{figure}[!htp]
\includegraphics[width=0.5\textwidth]{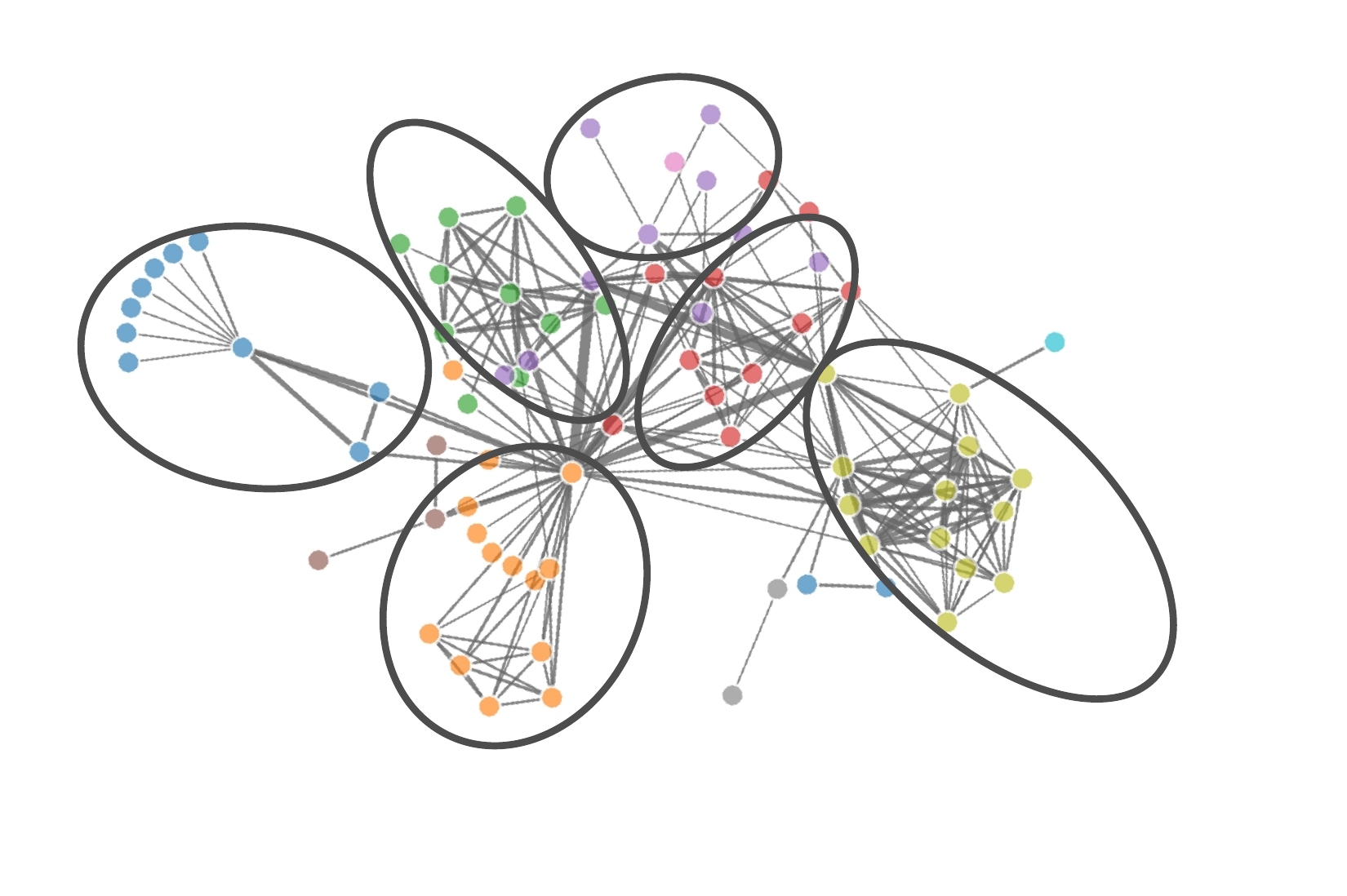}
\caption{An example of modular network along with its detected communities (encircled) is shown. The nodes are color-coded based on their memberships to these communities. }
\label{fig:modulillus}       % Give a unique label
\end{figure}

\clearpage
\noindent
One useful summary metric is the modularity, which is a measure of cluster formation in a network. More specifically,  the modularity $Q$ of a given network $A$ is defined with respect to a given grouping of its nodes.  We follow  \cite{fortunato2010community} where the grouping of the nodes is determined by a stochastic procedure  that reveals densely connected subgraphs. An illustration of a grouping is given in Figure \ref{fig:modulillus}. 
 
\begin{figure}[!htp]
\includegraphics[width=0.35\textwidth]{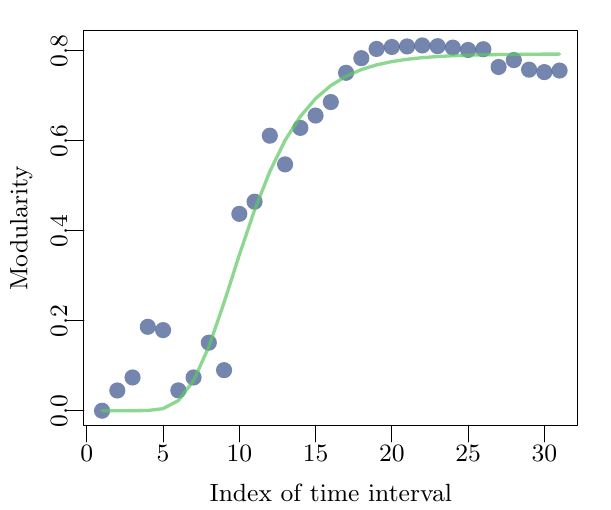}
\caption{The modularity of the network $A^{(n)(t)}$ is shown as a function of time $t = 20 \cdot n$, with $n=1 \ldots 35$. It is measure of the quality of the division of a graph into subgraphs.  }
\label{fig:modul}       % Give a unique label
\end{figure}

\begin{figure}[!htp]
\includegraphics[width=0.35\textwidth]{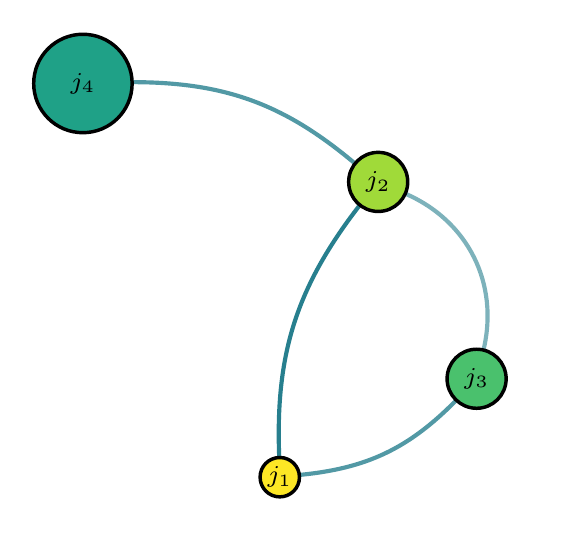}
\caption{ Example of a network with $C=1/3$ and $l = 8/6$ is shown. It has three triplets, one of which is closed (triangle). The lengths of the paths from $j_2$ to all others is 1, while that from $j_1$ to $j_4$ is 2.}
\label{fig:clusterIllu}       % Give a unique label
\end{figure}

\noindent
 Given this group membership, the modularity of $A$ is then computed according to the formula:
\begin{equation*}\label{modularity}
Q(A)=\frac{1}{2h} \sum_{{i,j}} \delta(c_i,c_j)(A_{ij}-k_i k_j/(2h) ) ,
\end{equation*}
where $h$ denotes the total number of edges, $k_i$ and $k_j$ are the degree of  nodes $i$ and $j$ respectively, $c_i$  labels the group to which   $i$ belongs, and $\delta$ is the Kronecker delta.   Figure \ref{fig:modul} shows the modularities  of the 35 matrices  $A^{(35)}(t)$ at  days $t=20 \cdot n$ for $n=1, \cdots, 35$.  The superscript $n$ in the notation $A^{(n)}(t)$, as mentioned above, indicates that $A$ is estimated  using the counts of the  $20\cdot n$ days of the study.  One can see   a jump in modularity on the tenth 20-day time interval, which we will  denote by $I_c$. This behavior may signal the  onset of an emerging power-law \cite{grindrod2018high}. We investigate if this is the case
by analyzing  two additional topological measures for the networks: mainly,  the clustering coefficient and the average path length \cite{albert2002statistical,newman2003structure}. These are generally used to classify networks into random, scale-free, or regular.
% Indeed,  we will show that their time evolution  does reveal an interesting change in the network topology. 
The clustering coefficient $C$ of a network  is a measure of transitivity that counts the ratio of the number closed triplets to the number of all (closed and open) triplets.  A triplet is  closed if all the three connections between  the three nodes exist and is open if one of the links is missing. The average path length $l$ of a network is given by the mean distance over all pairs of vertices, where distance is the number of edges in the shortest path joining them. 
 An illustration is shown in Figure \ref{fig:clusterIllu}. 

\begin{figure}[!htp] %
         \includegraphics[scale=1.3]{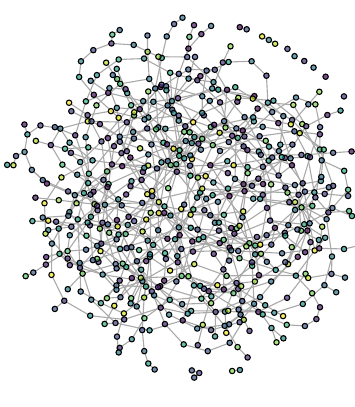} 
          \includegraphics[scale=1.3]{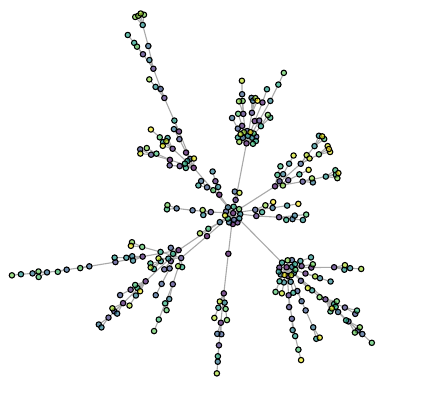}
          \includegraphics[scale=1.3]{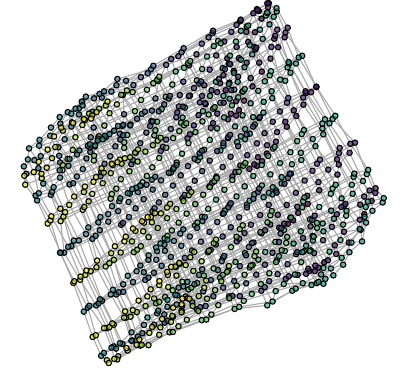}
         \caption{Examples of random, scale-free, and ordered (lattice-like) networks are shown respectively. }
\label{fig:randscalefree} 
\end{figure}

Small-world or scale-free networks (that is,  networks with node degrees and strengths  distributed according to a power-law) are characterized by  high clustering coefficients and low average path lengths compared with those of regular/ordered graphs \cite{newman2003structure,albert2002statistical}. Random graphs are, on the other hand, characterized by low average path lengths and low clustering coefficients compared to regular graphs. An illustration of the three different network types is shown in Figure \ref{fig:randscalefree}.

\begin{figure}[!htp]
\centering
\begin{subfigure}
{\includegraphics[width=0.45\textwidth]{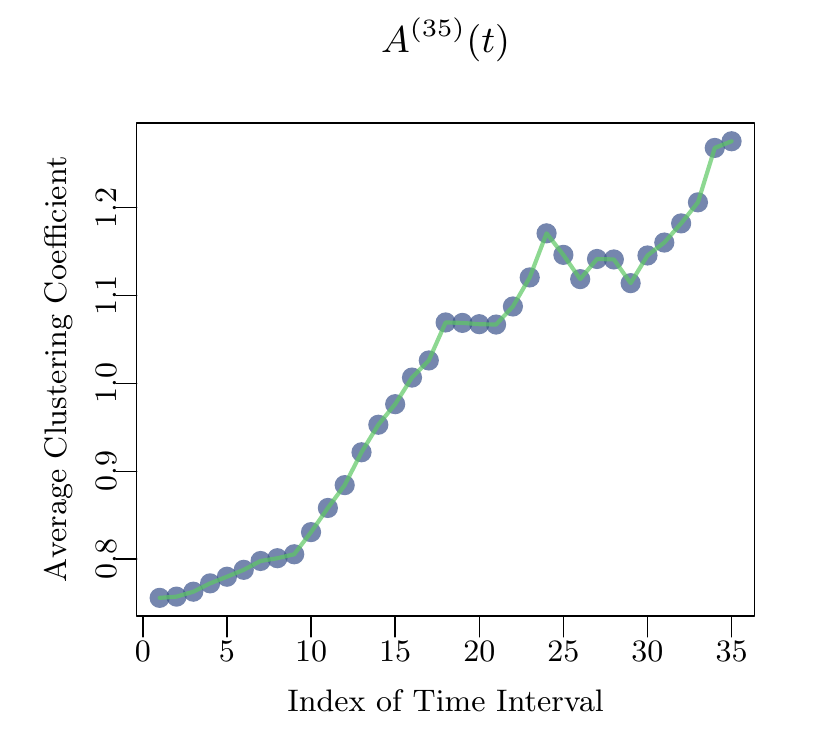} }
\end{subfigure}
\begin{subfigure}
{\includegraphics[width=0.43\textwidth]{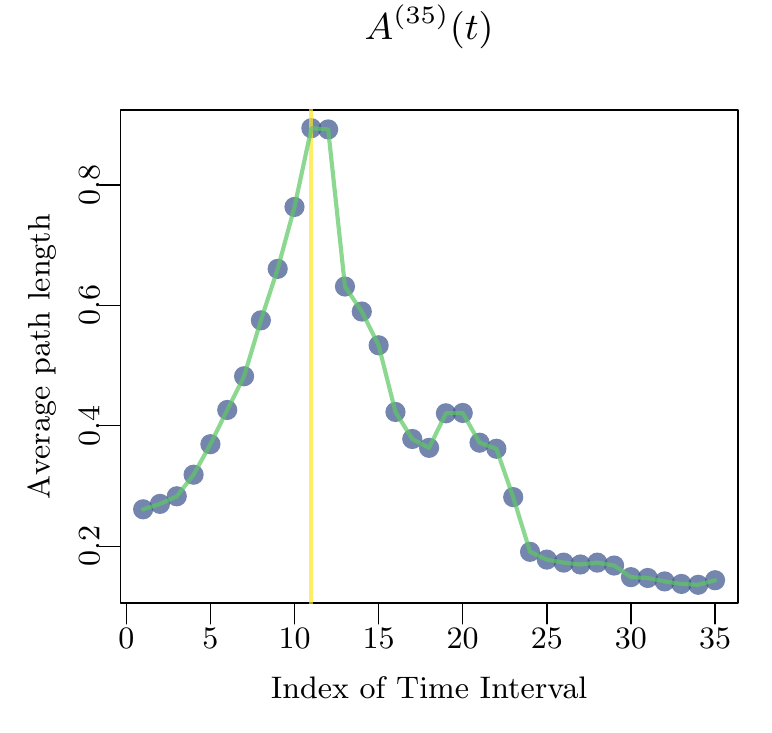}}%\label{fig:avpl}
\end{subfigure}
\caption{The figure shows the clustering coefficient and average path length for $A^{(35)}(t)$, which are the matrices estimated using the data from the start of the pandemic to the $700$-th day  ($35 \cdot 20$) evaluated at $t = 20 \cdot n$, where $n$ is the index of the time intervals.}
    \label{fig:cc}
\end{figure}

Figure \ref{fig:cc} shows  the clustering coefficients and the average paths lengths  for   the  matrices $A^{(35)}(t)$. Similar behavior of both $C$ and $l$ was observed for all $A^{(n)}$, with $n \geq 10$.
\noindent
The evolution of both $C$ and $l$ gives additional evidence for a transition at a point $I_c$. The clustering coefficient starts suddenly to increase. At the $10$-th interval  there is an abrupt jump in the average path length as well at $I_c$ (Figure \ref{fig:cc}). This is an indication of scale-freeness of the network. This property expedites the spread of epidemics unlike what would occur in ordered networks,  which are characterized by a slower spread because they possess  a high  $C$ and an $l$ that scales with system size \cite{pastor2015epidemic,newman2002spread}.

\begin{figure}[!htp]
\includegraphics[width=1\textwidth]{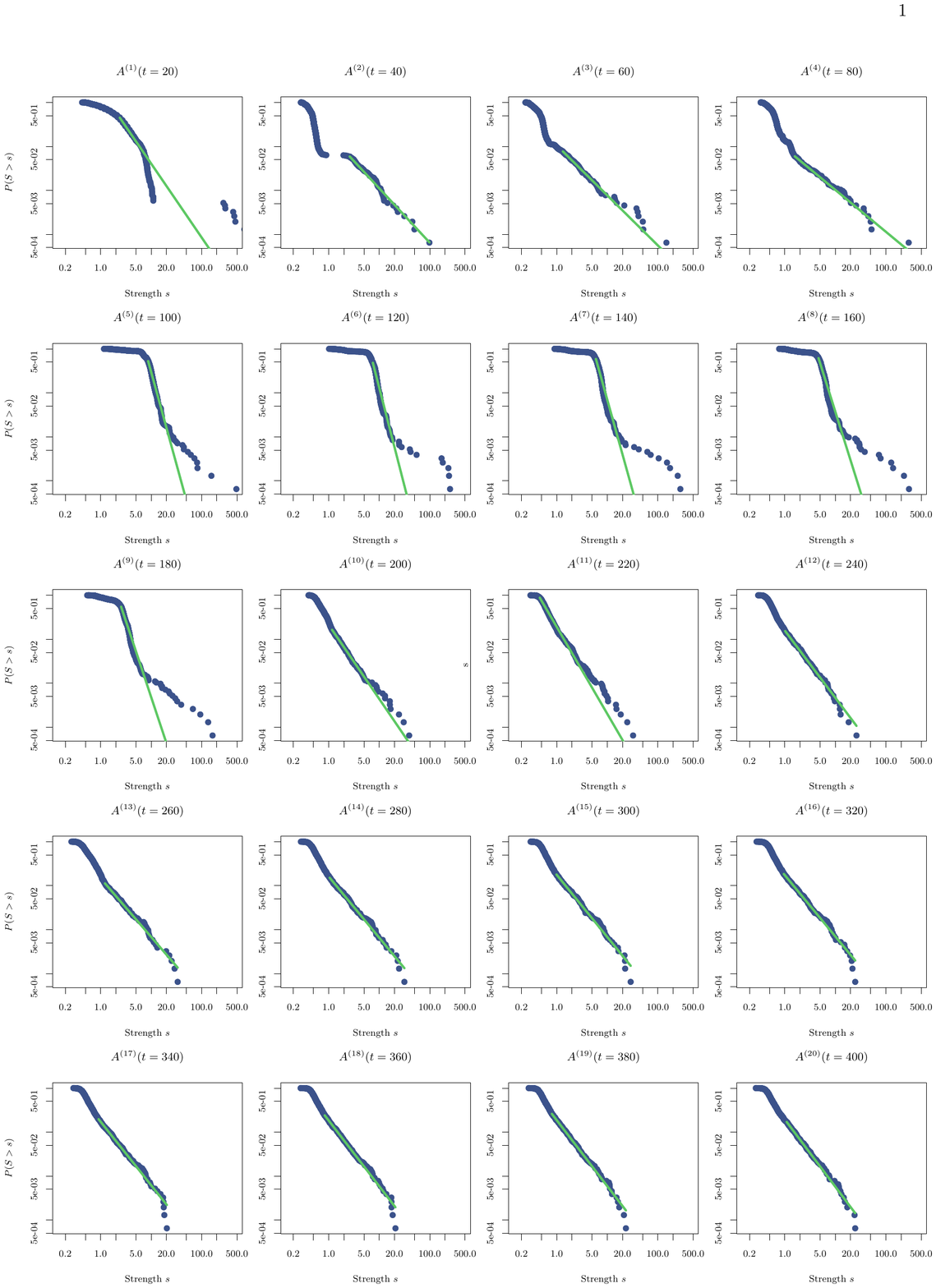}
\caption{Empirical and estimated distributions of the strength for the matrices $A^{(n)}(t = 20\cdot n)$ with $n=1, \ldots, 20$ are shown.} % Plot of $P(S>s)$  versus $s$ (strength of the nodes) in the log-log scale.   }
\label{fig:strengthdis}       % Give a unique label
\end{figure}

\begin{figure}[!htp]
\includegraphics[width=1\textwidth]{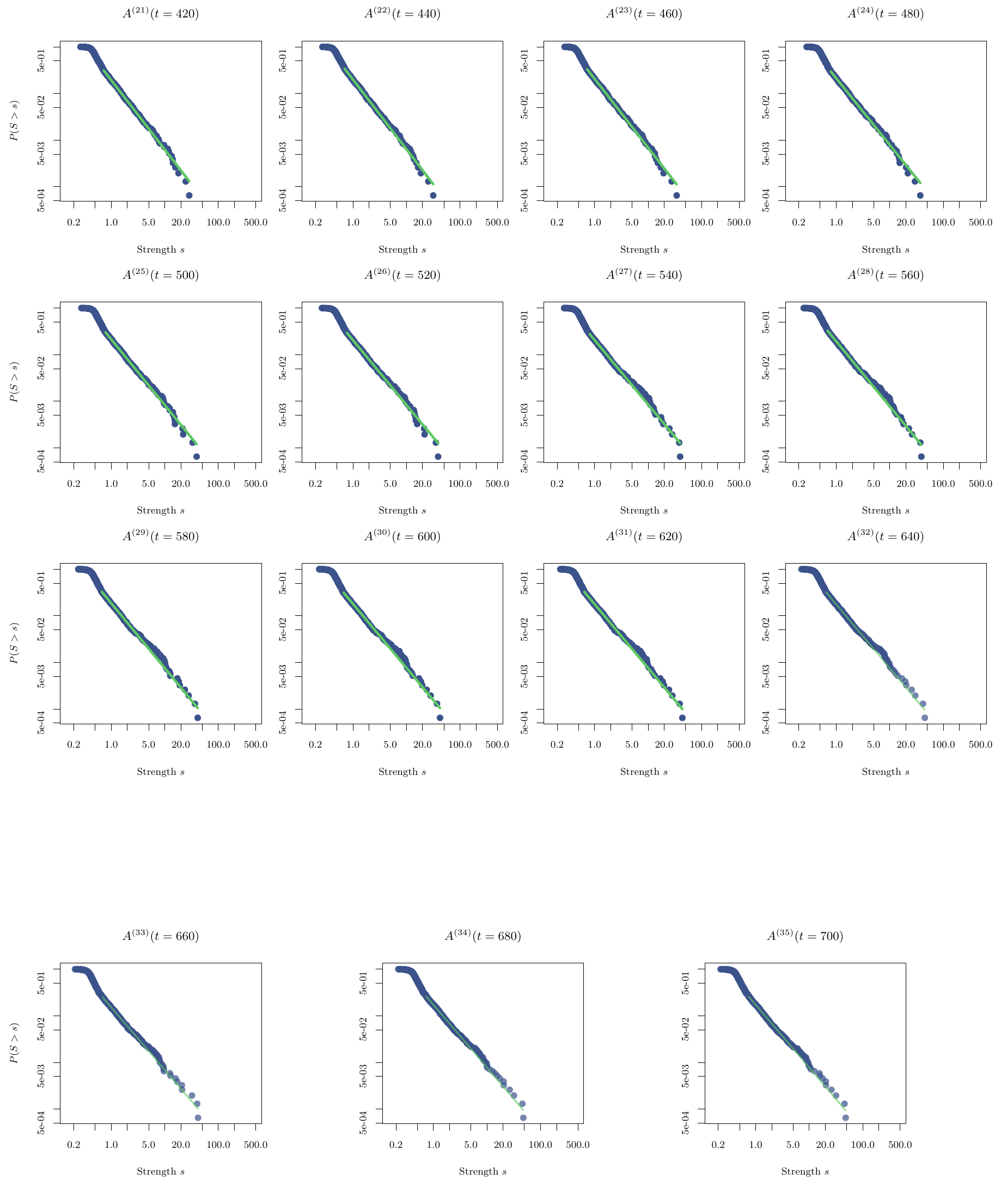}
\caption{Empirical and estimated distributions of the strength for the matrices $A^{(n)}(t = 20\cdot n)$ with $n=21, \ldots, 35$ are shown.}
\label{fig:strengthdis2}       % Give a unique label
\end{figure}

To characterize the transition to scale-freeness, we now analyse the distribution of the strengths of the nodes, as additional evidence for  change in the network topology at the $10$-th interval $I_C$.  A node's strength is the sum of the weights of its edges. Namely, for the $i$-th node:
\begin{equation*}\label{strength}
s_i = \sum_j A_{ij}
\end{equation*}
Figures \ref{fig:strengthdis} and \ref{fig:strengthdis2} show the empirical and estimated distributions of the strengths (in fact, the survival function $P(S> s)$) of $A^{(n)}(t = 20\cdot n)$, at the time intervals $n=1, \ldots, 35$ on a log-log scale. We note that a transition occurs at $t=20 \cdot 10$, where the distribution becomes linear, which is indicative of a power-law (Pareto distribution):  $P(X \leq x)=1-(\beta/x)^{\alpha-1}$, for $x\geq  \beta$. The exponent  $\alpha$  and the boundary value (scale)  $\beta$ are estimated by maximum likelihood following \cite{clauset2009power}. 
\clearpage

\begin{figure}[!htp]
\includegraphics[width=0.4\textwidth]{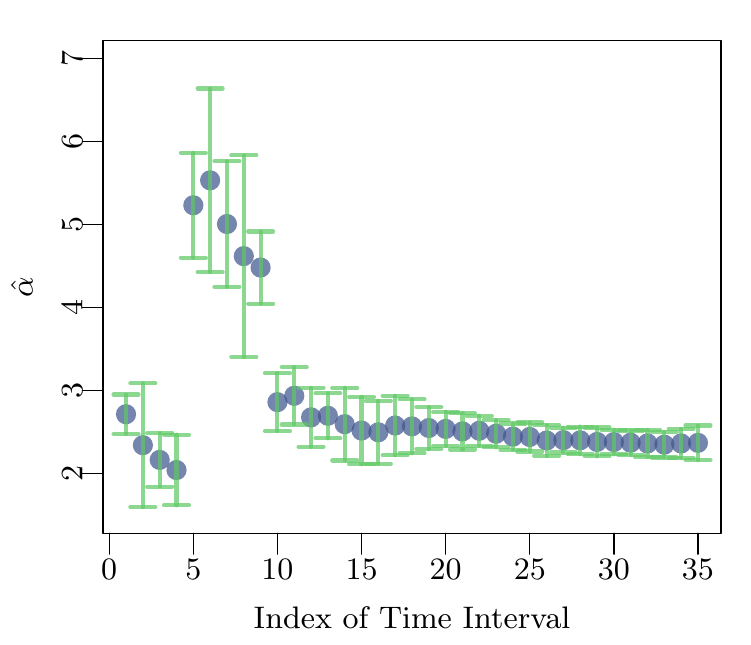}
\caption{Estimated  values $\hat{\alpha}$ of the exponent of the power-law distributions of the strengths
of the nodes of the 35 matrices  $A^{(n)}(t = 20\cdot n)$, $n=1, \ldots, 35$, which are represented in Figure \ref{fig:strengthdis}. Vertical bars indicate $\pm 2 \hat{\sigma}_{\hat{\alpha}}$.
}
\label{fig:bootstrapped}       % Give a unique label
\end{figure}

\clearpage
Figure  \ref{fig:bootstrapped} summarizes the estimates of the exponents for these networks and their standard errors  (obtained by non-parametric bootstrap).
After the 180th day, that is for time intervals labeled by the index $n \geq 10$, most  power-laws  have  very close  exponents of about 2.5. This  signals the stabilization of the network topology.  Thus, $I_C$ marks the onset of the emergence of the steady state network. We think that only above this point  any prescription of control measures is likely to be efficient as the revealed network topology, relying on the daily counts, has stabilized. 
One can wonder if there is any explanation on why the stable phase has set in during this interval $I_C$, and not before or after it.  $I_C$ chronologically coincides with the period between  August, 19, 2020,  and September 7, 2020.   
% We do not know the answer but are tempted to  advance some speculations. It may be suggestive to observe that the change to fixed-point testing sites
% (namely the availability  of sites where free testing is administered on some specific days in the week) occurred during this time. The increased number of testing has resulted in a higher number of observed cases, which will have affected the weights of the matrix $A$.
Perhaps, the blast in Beirut which occurred on August 4th and in the following weeks of social protests,  personal precaution measures (such as social  distancing and wearing of masks) were compromised. Either of these occurrences may have contributed to  the detected change in the network type. See the Appendix for the chronology. 

\section{Putting the analysis into action: Control measures}
\begin{figure}[!htp]
\includegraphics[width=0.6\textwidth]{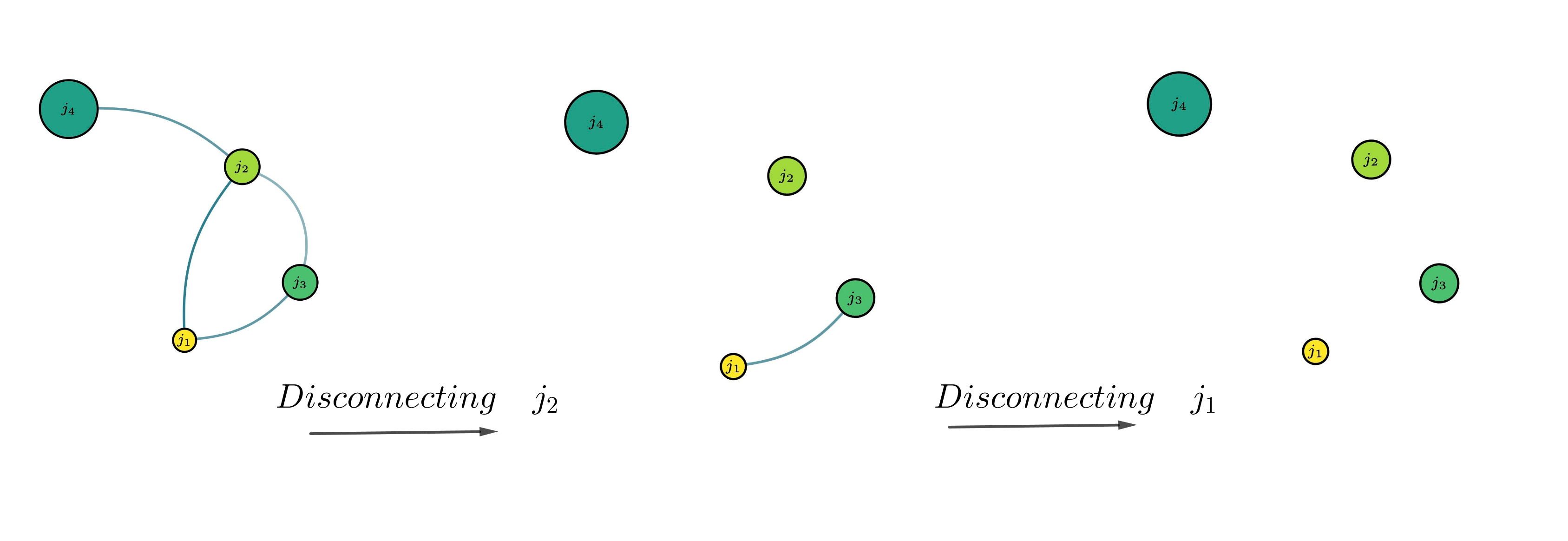}
\caption{The figure illustrates the iterative scheme. First, the node with the highest centrality $j_2$ is disconnected by removing all its links, as it is the node with the highest number of shortest paths. The resulting network has $j_1$ and $j_3$ with equal centralities and either one can be disconnected. This leads to a total loss of connectivity in the network at the end of the process.  }
\label{fig:removal}       % Give a unique label
\end{figure}

\noindent
Having fully characterized the network and identified the steady-state, we  now turn to a possible use of this analysis to guide  an optimal strategy for disease control. The strategy will identify some localities as candidates for being isolated or for having their connections to other localities curtailed.  The measure on which the identification is based is that of centrality of a node.
The betweenness centrality  of a node $v$ is defined as \cite{newman2003structure}:
\begin{equation*}\label{centrality}
g(v)=\sum _{{i\neq v\neq j}}{\frac  {\sigma _{{ij}}(v)}{\sigma _{{ij}}}}
\end{equation*}
where $\sigma_{ij}$ is the total number of shortest paths from node ${\displaystyle i}$ to node ${\displaystyle j}$ passing through $\sigma_{ij}(v)$. 
Therefore, the more central the node is, the more its removal has an effect on the network's connectivity, since its removal would yield a network with more disconnected subgraphs. 
The control strategy we propose involves an iterative procedure, where at each step the centralities of the nodes are computed, the node with the resulting highest centrality is removed, and the matrix $A$ is updated, as illustrated in Figure \ref{fig:removal}.
 
\begin{figure}[!htp]
\includegraphics[width=0.5\textwidth]{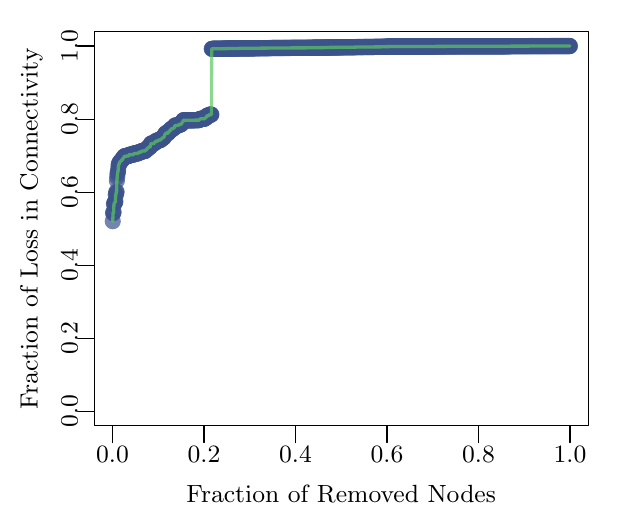}
\caption{The loss of connectivity versus the fraction of removed nodes for the cascading and non-cascading strategies. }
\label{fig:cascadenoncascade}       % Give a unique label
\end{figure}

\begin{figure}[!htp]
\includegraphics[width=0.5\textwidth]{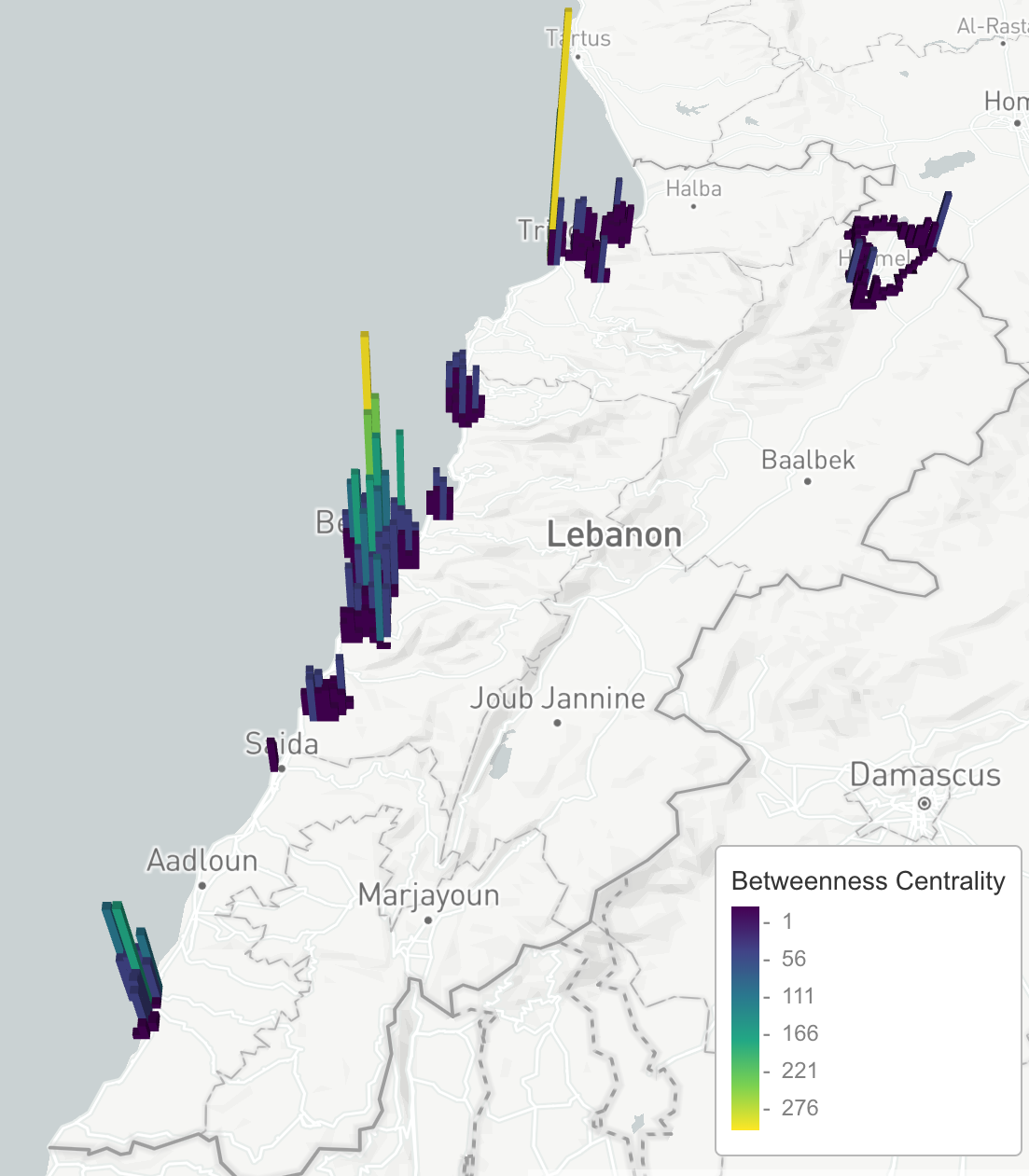}
\caption{The map shows the localities with the highest centrality whose removals lead to $80\%$ loss of connectivity.  }
\label{fig:cenonmap}       % Give a unique label
\end{figure}

\begin{figure}[!ht]
   \begin{minipage}{\textwidth}
   %  \centering
     \includegraphics[width=16cm, height=12cm]{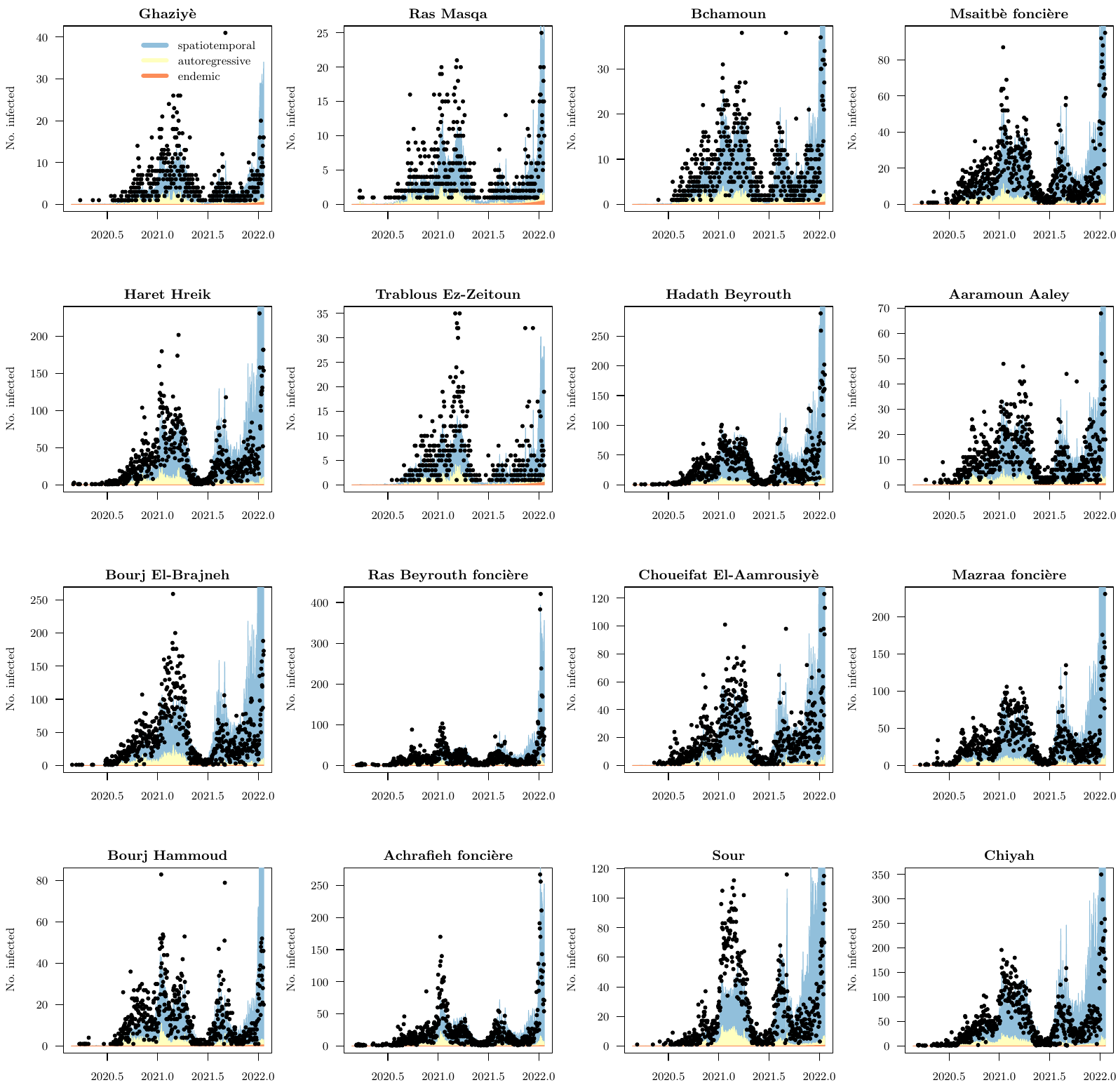} \clearpage%\quad % \hfill \hspace{-1.4cm}
    % \includegraphics[width=18cm, height=16cm]{localitylevePred2}\clearpage%\quad% \hfill \hspace{-1.4cm}
    % \includegraphics[width=18cm, height=16cm]{localitylevePred3}%\quad %\hfill \hspace{-1.4cm}
   %  \subcaption{First subfigure.}
      \end{minipage}\\[1em]
         \caption{The figure shows the counts, along with the fitted model, for twelve localities ordered by decreasing betweenness centrality. }\label{loclevel1}
      \end{figure}

Other removal schemes  of nodes in network exist, but the one we have just described has been suggested to incur the highest loss of connectivity for scale-free networks \cite{albert2000error,dong2013robustness,valdez2020cascading,buldyrev2010catastrophic,albert2004structural,edsberg2021understanding}.  In practice, candidate targets for intervention the localities corresponding to nodes with higher centralities.
We notice that at the policy level this strategy based on our analysis was indeed adopted.
The localities we have identified through this strategy were given priority in the national vaccination campaign. On the other hand, the recommendations we put forward based on this analysis were only partially adopted in targeting the high centrality localities for lockdown and intervention measures, as the decision making process involved other ministries and stakeholders. 
However, we conclude by considering theoretically  the would-be repercussions of such implementation. 
%have considered the theoretical repercussions  of such implementation. 
Clearly, the loss of connectivity would impede the evolution of the disease since the localities which are contributing the most to the infection would be isolated. For example,  removing  around $20\%$ of the most connected  localities on the basis of their betweenness centrality would lead to $80\%$ loss of connectivity as shown in Figure \ref{fig:cascadenoncascade}.   
Specifically, the localities causing $80\%$ loss of connectivity are shown in Figure \ref{fig:cenonmap},
while the fitted model of the top sixteen localities is shown in Figure \ref{loclevel1}. An animated map of the control strategy is available on this \href{https://www.dropbox.com/s/guhamz3p7op6b3y/animated.gif?dl=0}{hyperlink}.

\clearpage

\clearpage

\section{Conclusion}
In this paper, we have proposed a framework that can be used to inform control measures for epidemics in a country for which  infections counts aggregated over local
regions are available over time. In particular, we have  followed the evolution of the counts of COVID-19 cases in Lebanon at the level of local administrative units  at a daily resolution. 
The framework entails  fitting  an  auto-regressive model to the data; recovering  an underlying network over which the disease  propagates;  analyzing  such  time-evolving network to identify topological measures of node centrality that suggest an optimal control of the spread of the disease. Specifically, for the data about Covid-19 in Lebanon
% The parameters' variations across these intervals can be related to behavioral changes, for example changes in $\phi_{it}$, which is the contribution to infection from inter-locality interactions, reveal a change in commute patterns, whereas changes in $\nu_{it}$ are linked to changes in travel-related control measures, while changes in $\lambda_{it}$ can be associated to intra-locality daily activity and interaction rates. We have been able to recover the underlying network over which the disease was propagating.
the analysis of the topological metrics of the network has given us a hint into a transition to a steady state structure that governs  interactions between localities. After  identifying this steady state network, and characterizing it as a scale-free, we have proposed control measures based on betweenness centrality of its nodes.  The findings were taken into consideration in the national vaccination campaign for COVID-19, with  the identified  localities  given priority for vaccination. 
%\section*{Acknowledgement}

\section*{Appendix. Chronology of COVID-19 Pandemic  in Lebanon}
\noindent
% {\bf Period 1 (Feb 21- March 2020):} This was the first period of the outbreak in Lebanon, where on the 21st of February the first COVID-19 case was reported. Only suspected cases were referred for testing. 
% Suspected cases in this period were defined as:
% \begin{itemize}
% \item any person with symptoms and exposure (either contact of a confirmed case or a person arriving from countries with local transmission in the past 14 days)
% \item or any person requiring hospitalization due to respiratory infections. \\
% \end{itemize}
\noindent
In this appendix we summarize the chronology of the COVID-19 pandemic in Lebanon  from the first recorded case to February 2022. We divide this time interval into 4 periods, and  highlight 
the main governmental interventions taken to control the spread of the disease. 

Period 1 (February 2020 to June 2020).  The first cases are documented. Early lockdown measures are implemented with airport closure. Testing is carried out  for suspected cases, close contacts, and travelers. Cases are mainly within clusters. Aggressive contact tracing is adopted.

Period 2 (July 2020 to December 2020): The airport  reopens in July 2020. The  daily number of cases increases progressively and community transmission sets in. On August 4th, the Beirut blast occurs. 
% {\bf Period 2 (March 2020-April 2020):} In this period, the definition for a suspected case has been amended according to the below:
% \begin{itemize}
% \item asymptomatic contacts were tested
% \item any person with respiratory symptoms (even without exposure) was referred for testing.
% \end{itemize}
% Note: In the first two periods, suspected cases were referred to governmental hospitals for testing.
% \\

Period 3 (January 2021 to June 2021): The alpha  variant is  introduced. The case counts increase. Lockdown measures are implemented resulting in a  decrease of the recorded cases. However, after lockdown release, an increase of the number of infections is observed until  mid-March, with a  progressive and sustained decrease up to June.
% {\bf Period 3 (April 2020-August 2020):} In this period, Lebanon was in level 3 (cluster of cases) and each newly identified case necessitated mobilization of testing teams. The team goes to the exact building of the positive case where the whole building as well as the very small neighborhood would be tested. This was applicable as well to large institutions with cluster of cases.
% In this period, Syrian ITSs across Lebanon were screened (with no identified positive cases) as well as Palestinian camps and gatherings were being screened on regular basis. 
% \\

Period 4 (July 2021 to December 2021).  Introduction of the delta  variant, which progressively  replaces  the alpha variant. Two waves of delta are observed: July-September and November-December.

 Period 5 (January 2022-February 2022).  Introduction of the omicron  variant. High transmissibility of the new variant leads to high daily case counts reaching 10,000 on 1st Feb 2022.

% \noindent
% {\bf Period 4 (August 2020 – to date 2021):} Fixed point testing sites had been created in each district and contacts as well a s symptomatic people (who can’t afford to pay out-of-pocket for the PCR test) are asked to go to the corresponding sites in each district, following a fixed weekly schedule.

\section*{References}
%\newpage

%%%%%%%%%%%%%%%%%%%%%%%%%%%%%%%%%%%%%%%%%%%%%%
%%                                          %%
%% Backmatter begins here                   %%
%%                                          %%
%%%%%%%%%%%%%%%%%%%%%%%%%%%%%%%%%%%%%%%%%%%%%%

\begin{backmatter}

\section*{Acknowledgements}%% if any
The authors acknowledge the support of Prof. Rima Habib from the AUB's Faculty of Public Health for her unfailing support and invaluable advice to physicists and mathematicians as they worked with public data as well as Dr. Chadi Abdallah for helping in the early conception of the project. 
\section*{Funding}%% if any
The authors received no funding. 

%\section*{Availability of data and materials}%% if any
%Text for this section\ldots

%\section*{Ethics approval and consent to participate}%% if any
%Text for this section\ldots

\section*{Competing interests}
The authors declare that they have no competing interests.

%\section*{Consent for publication}%% if any
%Text for this section\ldots

\section*{Authors' contributions}
J.T contributed to the problem formulation and led the project, and together with S.N and S.M developed the model and numerical analysis involved in this manuscript. G.F and C.A worked on the geographical and the demographic data, R.H and H.S worked on the data processing of infection counts, N.G and H.H on the epidemiological analysis and the significance of the results and their implementation in the vaccination strategy.  
%Text for this section \ldots

%\section*{Authors' information}%% if any
%Text for this section\ldots

%%  After submission of the .TEX file,                     %%
%%  you will be prompted to submit your .BBL file.         %%

% if your bibliography is in bibtex format, use those commands:
\bibliographystyle{bmc-mathphys} % Style BST file (bmc-mathphys, vancouver, spbasic).
\bibliography{library}


\begin{thebibliography}{40}
\ifx \bisbn   \undefined \def \bisbn  #1{ISBN #1}\fi
\ifx \binits  \undefined \def \binits#1{#1}\fi
\ifx \bauthor  \undefined \def \bauthor#1{#1}\fi
\ifx \batitle  \undefined \def \batitle#1{#1}\fi
\ifx \bjtitle  \undefined \def \bjtitle#1{#1}\fi
\ifx \bvolume  \undefined \def \bvolume#1{\textbf{#1}}\fi
\ifx \byear  \undefined \def \byear#1{#1}\fi
\ifx \bissue  \undefined \def \bissue#1{#1}\fi
\ifx \bfpage  \undefined \def \bfpage#1{#1}\fi
\ifx \blpage  \undefined \def \blpage #1{#1}\fi
\ifx \burl  \undefined \def \burl#1{\textsf{#1}}\fi
\ifx \doiurl  \undefined \def \doiurl#1{\textsf{#1}}\fi
\ifx \betal  \undefined \def \betal{\textit{et al.}}\fi
\ifx \binstitute  \undefined \def \binstitute#1{#1}\fi
\ifx \binstitutionaled  \undefined \def \binstitutionaled#1{#1}\fi
\ifx \bctitle  \undefined \def \bctitle#1{#1}\fi
\ifx \beditor  \undefined \def \beditor#1{#1}\fi
\ifx \bpublisher  \undefined \def \bpublisher#1{#1}\fi
\ifx \bbtitle  \undefined \def \bbtitle#1{#1}\fi
\ifx \bedition  \undefined \def \bedition#1{#1}\fi
\ifx \bseriesno  \undefined \def \bseriesno#1{#1}\fi
\ifx \blocation  \undefined \def \blocation#1{#1}\fi
\ifx \bsertitle  \undefined \def \bsertitle#1{#1}\fi
\ifx \bsnm \undefined \def \bsnm#1{#1}\fi
\ifx \bsuffix \undefined \def \bsuffix#1{#1}\fi
\ifx \bparticle \undefined \def \bparticle#1{#1}\fi
\ifx \barticle \undefined \def \barticle#1{#1}\fi
\ifx \bconfdate \undefined \def \bconfdate #1{#1}\fi
\ifx \botherref \undefined \def \botherref #1{#1}\fi
\ifx \url \undefined \def \url#1{\textsf{#1}}\fi
\ifx \bchapter \undefined \def \bchapter#1{#1}\fi
\ifx \bbook \undefined \def \bbook#1{#1}\fi
\ifx \bcomment \undefined \def \bcomment#1{#1}\fi
\ifx \oauthor \undefined \def \oauthor#1{#1}\fi
\ifx \citeauthoryear \undefined \def \citeauthoryear#1{#1}\fi
\ifx \endbibitem  \undefined \def \endbibitem {}\fi
\ifx \bconflocation  \undefined \def \bconflocation#1{#1}\fi
\ifx \arxivurl  \undefined \def \arxivurl#1{\textsf{#1}}\fi
\csname PreBibitemsHook\endcsname

%%% 1
\bibitem{mohamadou2020review}
\begin{barticle}
\bauthor{\bsnm{Mohamadou}, \binits{Y.}},
\bauthor{\bsnm{Halidou}, \binits{A.}},
\bauthor{\bsnm{Kapen}, \binits{P.T.}}:
\batitle{A review of mathematical modeling, artificial intelligence and
  datasets used in the study, prediction and management of {COVID-19}}.
\bjtitle{Applied Intelligence}
\bvolume{50}(\bissue{11}),
\bfpage{3913}--\blpage{3925}
(\byear{2020})
\end{barticle}
\endbibitem

%%% 2
\bibitem{guan2020modeling}
\begin{barticle}
\bauthor{\bsnm{Guan}, \binits{J.}},
\bauthor{\bsnm{Wei}, \binits{Y.}},
\bauthor{\bsnm{Zhao}, \binits{Y.}},
\bauthor{\bsnm{Chen}, \binits{F.}}:
\batitle{Modeling the transmission dynamics of {COVID-19} epidemic: a
  systematic review}.
\bjtitle{Journal of Biomedical Research}
\bvolume{34}(\bissue{6}),
\bfpage{422}
(\byear{2020})
\end{barticle}
\endbibitem

%%% 3
\bibitem{celani2021endemic}
\begin{botherref}
\oauthor{\bsnm{Celani}, \binits{A.}},
\oauthor{\bsnm{Giudici}, \binits{P.}}:
Endemic--epidemic models to understand {COVID-19} spatio-temporal evolution.
Spatial Statistics,
100528
(2021)
\end{botherref}
\endbibitem

%%% 4
\bibitem{ssentongo2021pan}
\begin{botherref}
\oauthor{\bsnm{Ssentongo}, \binits{P.}},
\oauthor{\bsnm{Fronterre}, \binits{C.}},
\oauthor{\bsnm{Geronimo}, \binits{A.}},
\oauthor{\bsnm{Greybush}, \binits{S.J.}},
\oauthor{\bsnm{Mbabazi}, \binits{P.K.}},
\oauthor{\bsnm{Muvawala}, \binits{J.}},
\oauthor{\bsnm{Nahalamba}, \binits{S.B.}},
\oauthor{\bsnm{Omadi}, \binits{P.O.}},
\oauthor{\bsnm{Opar}, \binits{B.T.}},
\oauthor{\bsnm{Sinnar}, \binits{S.A.}}, et al.:
Pan-{A}frican evolution of within-and between-country {COVID-19} dynamics.
Proceedings of the National Academy of Sciences
\textbf{118}(28)
(2021)
\end{botherref}
\endbibitem

%%% 5
\bibitem{dickson2020assessing}
\begin{barticle}
\bauthor{\bsnm{Dickson}, \binits{M.M.}},
\bauthor{\bsnm{Espa}, \binits{G.}},
\bauthor{\bsnm{Giuliani}, \binits{D.}},
\bauthor{\bsnm{Santi}, \binits{F.}},
\bauthor{\bsnm{Savadori}, \binits{L.}}:
\batitle{Assessing the effect of containment measures on the spatio-temporal
  dynamic of {COVID-19} in {I}taly}.
\bjtitle{Nonlinear Dynamics}
\bvolume{101}(\bissue{3}),
\bfpage{1833}--\blpage{1846}
(\byear{2020})
\end{barticle}
\endbibitem

%%% 6
\bibitem{gozzi2021importance}
\begin{botherref}
\oauthor{\bsnm{Gozzi}, \binits{N.}},
\oauthor{\bsnm{Bajardi}, \binits{P.}},
\oauthor{\bsnm{Perra}, \binits{N.}}:
The importance of non-pharmaceutical interventions during the {COVID-19}
  vaccine rollout.
medRxiv
(2021)
\end{botherref}
\endbibitem

%%% 7
\bibitem{perra2021non}
\begin{botherref}
\oauthor{\bsnm{Perra}, \binits{N.}}:
Non-pharmaceutical interventions during the {COVID-19} pandemic: A review.
Physics Reports
(2021)
\end{botherref}
\endbibitem

%%% 8
\bibitem{brockmann2013hidden}
\begin{barticle}
\bauthor{\bsnm{Brockmann}, \binits{D.}},
\bauthor{\bsnm{Helbing}, \binits{D.}}:
\batitle{The hidden geometry of complex, network-driven contagion phenomena}.
\bjtitle{science}
\bvolume{342}(\bissue{6164}),
\bfpage{1337}--\blpage{1342}
(\byear{2013})
\end{barticle}
\endbibitem

%%% 9
\bibitem{schlosser2021finding}
\begin{barticle}
\bauthor{\bsnm{Schlosser}, \binits{F.}},
\bauthor{\bsnm{Brockmann}, \binits{D.}}:
\batitle{Finding disease outbreak locations from human mobility data}.
\bjtitle{EPJ data science}
\bvolume{10}(\bissue{1}),
\bfpage{52}
(\byear{2021})
\end{barticle}
\endbibitem

%%% 10
\bibitem{zhu2021high}
\begin{barticle}
\bauthor{\bsnm{Zhu}, \binits{S.}},
\bauthor{\bsnm{Bukharin}, \binits{A.}},
\bauthor{\bsnm{Xie}, \binits{L.}},
\bauthor{\bsnm{Santillana}, \binits{M.}},
\bauthor{\bsnm{Yang}, \binits{S.}},
\bauthor{\bsnm{Xie}, \binits{Y.}}:
\batitle{High-resolution spatio-temporal model for county-level {COVID-19}
  activity in the {US}}.
\bjtitle{ACM Transactions on Management Information Systems (TMIS)}
\bvolume{12}(\bissue{4}),
\bfpage{1}--\blpage{20}
(\byear{2021})
\end{barticle}
\endbibitem

%%% 11
\bibitem{chiang2021hawkes}
\begin{botherref}
\oauthor{\bsnm{Chiang}, \binits{W.-H.}},
\oauthor{\bsnm{Liu}, \binits{X.}},
\oauthor{\bsnm{Mohler}, \binits{G.}}:
Hawkes process modeling of {COVID-19} with mobility leading indicators and
  spatial covariates.
International journal of forecasting
(2021)
\end{botherref}
\endbibitem

%%% 12
\bibitem{giudici2021network}
\begin{botherref}
\oauthor{\bsnm{Giudici}, \binits{P.}},
\oauthor{\bsnm{Pagnottoni}, \binits{P.}},
\oauthor{\bsnm{Spelta}, \binits{A.}}:
Network self-exciting point processes to measure health impacts of {COVID-19}.
Available at SSRN 3892998
(2021)
\end{botherref}
\endbibitem

%%% 13
\bibitem{hhh}
\begin{barticle}
\bauthor{\bsnm{Held}, \binits{L.}},
\bauthor{\bsnm{Höhle}, \binits{M.}},
\bauthor{\bsnm{Hofmann}, \binits{M.}}:
\batitle{A statistical framework for the analysis of multivariate infectious
  disease surveillance counts}.
\bjtitle{Statistical Modelling}
\bvolume{5},
\bfpage{187}--\blpage{199}
(\byear{2005})
\end{barticle}
\endbibitem

%%% 14
\bibitem{meyer2014spatio}
\begin{botherref}
\oauthor{\bsnm{Meyer}, \binits{S.}},
\oauthor{\bsnm{Held}, \binits{L.}},
\oauthor{\bsnm{H{\"o}hle}, \binits{M.}}:
Spatio-temporal analysis of epidemic phenomena using the {R} package
  surveillance.
arXiv preprint arXiv:1411.0416
(2014)
\end{botherref}
\endbibitem

%%% 15
\bibitem{shandilya2011inferring}
\begin{barticle}
\bauthor{\bsnm{Shandilya}, \binits{S.G.}},
\bauthor{\bsnm{Timme}, \binits{M.}}:
\batitle{Inferring network topology from complex dynamics}.
\bjtitle{New Journal of Physics}
\bvolume{13}(\bissue{1}),
\bfpage{013004}
(\byear{2011})
\end{barticle}
\endbibitem

%%% 16
\bibitem{prabakaran2014paradoxical}
\begin{barticle}
\bauthor{\bsnm{Prabakaran}, \binits{S.}},
\bauthor{\bsnm{Gunawardena}, \binits{J.}},
\bauthor{\bsnm{Sontag}, \binits{E.}}:
\batitle{Paradoxical results in perturbation-based signaling network
  reconstruction}.
\bjtitle{Biophysical journal}
\bvolume{106}(\bissue{12}),
\bfpage{2720}--\blpage{2728}
(\byear{2014})
\end{barticle}
\endbibitem

%%% 17
\bibitem{yu2010estimating}
\begin{barticle}
\bauthor{\bsnm{Yu}, \binits{D.}}:
\batitle{Estimating the topology of complex dynamical networks by steady state
  control: Generality and limitation}.
\bjtitle{Automatica}
\bvolume{46}(\bissue{12}),
\bfpage{2035}--\blpage{2040}
(\byear{2010})
\end{barticle}
\endbibitem

%%% 18
\bibitem{yu2010inferring}
\begin{barticle}
\bauthor{\bsnm{Yu}, \binits{D.}},
\bauthor{\bsnm{Parlitz}, \binits{U.}}:
\batitle{Inferring local dynamics and connectivity of spatially extended
  systems with long-range links based on steady-state stabilization}.
\bjtitle{Physical Review E}
\bvolume{82}(\bissue{2}),
\bfpage{026108}
(\byear{2010})
\end{barticle}
\endbibitem

%%% 19
\bibitem{wan2014inferring}
\begin{barticle}
\bauthor{\bsnm{Wan}, \binits{X.}},
\bauthor{\bsnm{Liu}, \binits{J.}},
\bauthor{\bsnm{Cheung}, \binits{W.K.}},
\bauthor{\bsnm{Tong}, \binits{T.}}:
\batitle{Inferring epidemic network topology from surveillance data}.
\bjtitle{PLoS One}
\bvolume{9}(\bissue{6}),
\bfpage{100661}
(\byear{2014})
\end{barticle}
\endbibitem

%%% 20
\bibitem{pajevic2009efficient}
\begin{barticle}
\bauthor{\bsnm{Pajevic}, \binits{S.}},
\bauthor{\bsnm{Plenz}, \binits{D.}}:
\batitle{Efficient network reconstruction from dynamical cascades identifies
  small-world topology of neuronal avalanches}.
\bjtitle{PLoS computational biology}
\bvolume{5}(\bissue{1}),
\bfpage{1000271}
(\byear{2009})
\end{barticle}
\endbibitem

%%% 21
\bibitem{braunstein2019network}
\begin{barticle}
\bauthor{\bsnm{Braunstein}, \binits{A.}},
\bauthor{\bsnm{Ingrosso}, \binits{A.}},
\bauthor{\bsnm{Muntoni}, \binits{A.P.}}:
\batitle{Network reconstruction from infection cascades}.
\bjtitle{Journal of the Royal Society Interface}
\bvolume{16}(\bissue{151}),
\bfpage{20180844}
(\byear{2019})
\end{barticle}
\endbibitem

%%% 22
\bibitem{meyer2014power}
\begin{barticle}
\bauthor{\bsnm{Meyer}, \binits{S.}},
\bauthor{\bsnm{Held}, \binits{L.}}:
\batitle{Power-law models for infectious disease spread}.
\bjtitle{The Annals of Applied Statistics}
\bvolume{8}(\bissue{3}),
\bfpage{1612}--\blpage{1639}
(\byear{2014})
\end{barticle}
\endbibitem

%%% 23
\bibitem{zipf1946p}
\begin{barticle}
\bauthor{\bsnm{Zipf}, \binits{G.K.}}:
\batitle{The p 1 p 2/d hypothesis: on the intercity movement of persons}.
\bjtitle{American sociological review}
\bvolume{11}(\bissue{6}),
\bfpage{677}--\blpage{686}
(\byear{1946})
\end{barticle}
\endbibitem

%%% 24
\bibitem{simini2012universal}
\begin{barticle}
\bauthor{\bsnm{Simini}, \binits{F.}},
\bauthor{\bsnm{Gonz{\'a}lez}, \binits{M.C.}},
\bauthor{\bsnm{Maritan}, \binits{A.}},
\bauthor{\bsnm{Barab{\'a}si}, \binits{A.-L.}}:
\batitle{A universal model for mobility and migration patterns}.
\bjtitle{Nature}
\bvolume{484}(\bissue{7392}),
\bfpage{96}--\blpage{100}
(\byear{2012})
\end{barticle}
\endbibitem

%%% 25
\bibitem{barthelemy2011spatial}
\begin{barticle}
\bauthor{\bsnm{Barth{\'e}lemy}, \binits{M.}}:
\batitle{Spatial networks}.
\bjtitle{Physics Reports}
\bvolume{499}(\bissue{1-3}),
\bfpage{1}--\blpage{101}
(\byear{2011})
\end{barticle}
\endbibitem

%%% 26
\bibitem{verdeil2007atlas}
\begin{bbook}
\bauthor{\bsnm{Verdeil}, \binits{E.}},
\bauthor{\bsnm{Faour}, \binits{G.}},
\bauthor{\bsnm{Velut}, \binits{S.}},
\bauthor{\bsnm{Hamz{\'e}}, \binits{M.}},
\bauthor{\bsnm{Mermier}, \binits{F.}}:
\bbtitle{Atlas du {LIBAN}}.
\bpublisher{Presses de l'Ifpo}, \blocation{???}
(\byear{2007})
\end{bbook}
\endbibitem

%%% 27
\bibitem{karumanagoundar2021secondary}
\begin{barticle}
\bauthor{\bsnm{Karumanagoundar}, \binits{K.}},
\bauthor{\bsnm{Raju}, \binits{M.}},
\bauthor{\bsnm{Ponnaiah}, \binits{M.}},
\bauthor{\bsnm{Kaur}, \binits{P.}},
\bauthor{\bsnm{Rubeshkumar}, \binits{P.}},
\bauthor{\bsnm{Sakthivel}, \binits{M.}},
\bauthor{\bsnm{Shanmugiah}, \binits{P.}},
\bauthor{\bsnm{Ganeshkumar}, \binits{P.}},
\bauthor{\bsnm{Muthusamy}, \binits{S.K.}},
\bauthor{\bsnm{Sendhilkumar}, \binits{M.}}, \betal:
\batitle{Secondary attack rate of {COVID-19} among contacts and risk factors,
  tamil nadu, march--may 2020: a retrospective cohort study}.
\bjtitle{BMJ open}
\bvolume{11}(\bissue{11}),
\bfpage{051491}
(\byear{2021})
\end{barticle}
\endbibitem

%%% 28
\bibitem{fortunato2010community}
\begin{barticle}
\bauthor{\bsnm{Fortunato}, \binits{S.}}:
\batitle{Community detection in graphs}.
\bjtitle{Physics reports}
\bvolume{486}(\bissue{3-5}),
\bfpage{75}--\blpage{174}
(\byear{2010})
\end{barticle}
\endbibitem

%%% 29
\bibitem{grindrod2018high}
\begin{barticle}
\bauthor{\bsnm{Grindrod}, \binits{P.}},
\bauthor{\bsnm{Higham}, \binits{D.J.}}:
\batitle{High modularity creates scaling laws}.
\bjtitle{Scientific reports}
\bvolume{8}(\bissue{1}),
\bfpage{1}--\blpage{9}
(\byear{2018})
\end{barticle}
\endbibitem

%%% 30
\bibitem{albert2002statistical}
\begin{barticle}
\bauthor{\bsnm{Albert}, \binits{R.}},
\bauthor{\bsnm{Barab{\'a}si}, \binits{A.-L.}}:
\batitle{Statistical mechanics of complex networks}.
\bjtitle{Reviews of modern physics}
\bvolume{74}(\bissue{1}),
\bfpage{47}
(\byear{2002})
\end{barticle}
\endbibitem

%%% 31
\bibitem{newman2003structure}
\begin{barticle}
\bauthor{\bsnm{Newman}, \binits{M.E.}}:
\batitle{The structure and function of complex networks}.
\bjtitle{SIAM review}
\bvolume{45}(\bissue{2}),
\bfpage{167}--\blpage{256}
(\byear{2003})
\end{barticle}
\endbibitem

%%% 32
\bibitem{pastor2015epidemic}
\begin{barticle}
\bauthor{\bsnm{Pastor-Satorras}, \binits{R.}},
\bauthor{\bsnm{Castellano}, \binits{C.}},
\bauthor{\bsnm{Van~Mieghem}, \binits{P.}},
\bauthor{\bsnm{Vespignani}, \binits{A.}}:
\batitle{Epidemic processes in complex networks}.
\bjtitle{Reviews of modern physics}
\bvolume{87}(\bissue{3}),
\bfpage{925}
(\byear{2015})
\end{barticle}
\endbibitem

%%% 33
\bibitem{newman2002spread}
\begin{barticle}
\bauthor{\bsnm{Newman}, \binits{M.E.}}:
\batitle{Spread of epidemic disease on networks}.
\bjtitle{Physical review E}
\bvolume{66}(\bissue{1}),
\bfpage{016128}
(\byear{2002})
\end{barticle}
\endbibitem

%%% 34
\bibitem{clauset2009power}
\begin{barticle}
\bauthor{\bsnm{Clauset}, \binits{A.}},
\bauthor{\bsnm{Shalizi}, \binits{C.R.}},
\bauthor{\bsnm{Newman}, \binits{M.E.}}:
\batitle{Power-law distributions in empirical data}.
\bjtitle{SIAM review}
\bvolume{51}(\bissue{4}),
\bfpage{661}--\blpage{703}
(\byear{2009})
\end{barticle}
\endbibitem

%%% 35
\bibitem{albert2000error}
\begin{barticle}
\bauthor{\bsnm{Albert}, \binits{R.}},
\bauthor{\bsnm{Jeong}, \binits{H.}},
\bauthor{\bsnm{Barab{\'a}si}, \binits{A.-L.}}:
\batitle{Error and attack tolerance of complex networks}.
\bjtitle{nature}
\bvolume{406}(\bissue{6794}),
\bfpage{378}--\blpage{382}
(\byear{2000})
\end{barticle}
\endbibitem

%%% 36
\bibitem{dong2013robustness}
\begin{barticle}
\bauthor{\bsnm{Dong}, \binits{G.}},
\bauthor{\bsnm{Gao}, \binits{J.}},
\bauthor{\bsnm{Du}, \binits{R.}},
\bauthor{\bsnm{Tian}, \binits{L.}},
\bauthor{\bsnm{Stanley}, \binits{H.E.}},
\bauthor{\bsnm{Havlin}, \binits{S.}}:
\batitle{Robustness of network of networks under targeted attack}.
\bjtitle{Physical Review E}
\bvolume{87}(\bissue{5}),
\bfpage{052804}
(\byear{2013})
\end{barticle}
\endbibitem

%%% 37
\bibitem{valdez2020cascading}
\begin{barticle}
\bauthor{\bsnm{Valdez}, \binits{L.D.}},
\bauthor{\bsnm{Shekhtman}, \binits{L.}},
\bauthor{\bsnm{La~Rocca}, \binits{C.E.}},
\bauthor{\bsnm{Zhang}, \binits{X.}},
\bauthor{\bsnm{Buldyrev}, \binits{S.V.}},
\bauthor{\bsnm{Trunfio}, \binits{P.A.}},
\bauthor{\bsnm{Braunstein}, \binits{L.A.}},
\bauthor{\bsnm{Havlin}, \binits{S.}}:
\batitle{Cascading failures in complex networks}.
\bjtitle{Journal of Complex Networks}
\bvolume{8}(\bissue{2}),
\bfpage{013}
(\byear{2020})
\end{barticle}
\endbibitem

%%% 38
\bibitem{buldyrev2010catastrophic}
\begin{barticle}
\bauthor{\bsnm{Buldyrev}, \binits{S.V.}},
\bauthor{\bsnm{Parshani}, \binits{R.}},
\bauthor{\bsnm{Paul}, \binits{G.}},
\bauthor{\bsnm{Stanley}, \binits{H.E.}},
\bauthor{\bsnm{Havlin}, \binits{S.}}:
\batitle{Catastrophic cascade of failures in interdependent networks}.
\bjtitle{Nature}
\bvolume{464}(\bissue{7291}),
\bfpage{1025}--\blpage{1028}
(\byear{2010})
\end{barticle}
\endbibitem

%%% 39
\bibitem{albert2004structural}
\begin{barticle}
\bauthor{\bsnm{Albert}, \binits{R.}},
\bauthor{\bsnm{Albert}, \binits{I.}},
\bauthor{\bsnm{Nakarado}, \binits{G.L.}}:
\batitle{Structural vulnerability of the {N}orth {A}merican power grid}.
\bjtitle{Physical review E}
\bvolume{69}(\bissue{2}),
\bfpage{025103}
(\byear{2004})
\end{barticle}
\endbibitem

%%% 40
\bibitem{edsberg2021understanding}
\begin{barticle}
\bauthor{\bsnm{Edsberg~M{\o}llgaard}, \binits{P.}},
\bauthor{\bsnm{Lehmann}, \binits{S.}},
\bauthor{\bsnm{Alessandretti}, \binits{L.}}:
\batitle{Understanding components of mobility during the {COVID-19} pandemic}.
\bjtitle{Philosophical Transactions of the Royal Society A}
\bvolume{380}(\bissue{2214}),
\bfpage{20210118}
(\byear{2021})
\end{barticle}
\endbibitem

\end{thebibliography}

\newcommand{\BMCxmlcomment}[1]{}

\BMCxmlcomment{

<refgrp>

<bibl id="B1">
  <title><p>A review of mathematical modeling, artificial intelligence and
  datasets used in the study, prediction and management of
  {COVID-19}</p></title>
  <aug>
    <au><snm>Mohamadou</snm><fnm>Y</fnm></au>
    <au><snm>Halidou</snm><fnm>A</fnm></au>
    <au><snm>Kapen</snm><fnm>PT</fnm></au>
  </aug>
  <source>Applied Intelligence</source>
  <publisher>Springer</publisher>
  <pubdate>2020</pubdate>
  <volume>50</volume>
  <issue>11</issue>
  <fpage>3913</fpage>
  <lpage>-3925</lpage>
</bibl>

<bibl id="B2">
  <title><p>Modeling the transmission dynamics of {COVID-19} epidemic: a
  systematic review</p></title>
  <aug>
    <au><snm>Guan</snm><fnm>J</fnm></au>
    <au><snm>Wei</snm><fnm>Y</fnm></au>
    <au><snm>Zhao</snm><fnm>Y</fnm></au>
    <au><snm>Chen</snm><fnm>F</fnm></au>
  </aug>
  <source>Journal of Biomedical Research</source>
  <publisher>Education Department of Jiangsu Province</publisher>
  <pubdate>2020</pubdate>
  <volume>34</volume>
  <issue>6</issue>
  <fpage>422</fpage>
</bibl>

<bibl id="B3">
  <title><p>Endemic--epidemic models to understand {COVID-19} spatio-temporal
  evolution</p></title>
  <aug>
    <au><snm>Celani</snm><fnm>A</fnm></au>
    <au><snm>Giudici</snm><fnm>P</fnm></au>
  </aug>
  <source>Spatial Statistics</source>
  <publisher>Elsevier</publisher>
  <pubdate>2021</pubdate>
  <fpage>100528</fpage>
</bibl>

<bibl id="B4">
  <title><p>Pan-{A}frican evolution of within-and between-country {COVID-19}
  dynamics</p></title>
  <aug>
    <au><snm>Ssentongo</snm><fnm>P</fnm></au>
    <au><snm>Fronterre</snm><fnm>C</fnm></au>
    <au><snm>Geronimo</snm><fnm>A</fnm></au>
    <au><snm>Greybush</snm><fnm>SJ</fnm></au>
    <au><snm>Mbabazi</snm><fnm>PK</fnm></au>
    <au><snm>Muvawala</snm><fnm>J</fnm></au>
    <au><snm>Nahalamba</snm><fnm>SB</fnm></au>
    <au><snm>Omadi</snm><fnm>PO</fnm></au>
    <au><snm>Opar</snm><fnm>BT</fnm></au>
    <au><snm>Sinnar</snm><fnm>SA</fnm></au>
    <au><cnm>others</cnm></au>
  </aug>
  <source>Proceedings of the National Academy of Sciences</source>
  <publisher>National Acad Sciences</publisher>
  <pubdate>2021</pubdate>
  <volume>118</volume>
  <issue>28</issue>
</bibl>

<bibl id="B5">
  <title><p>Assessing the effect of containment measures on the spatio-temporal
  dynamic of {COVID-19} in {I}taly</p></title>
  <aug>
    <au><snm>Dickson</snm><fnm>MM</fnm></au>
    <au><snm>Espa</snm><fnm>G</fnm></au>
    <au><snm>Giuliani</snm><fnm>D</fnm></au>
    <au><snm>Santi</snm><fnm>F</fnm></au>
    <au><snm>Savadori</snm><fnm>L</fnm></au>
  </aug>
  <source>Nonlinear Dynamics</source>
  <publisher>Springer</publisher>
  <pubdate>2020</pubdate>
  <volume>101</volume>
  <issue>3</issue>
  <fpage>1833</fpage>
  <lpage>-1846</lpage>
</bibl>

<bibl id="B6">
  <title><p>The importance of non-pharmaceutical interventions during the
  {COVID-19} vaccine rollout</p></title>
  <aug>
    <au><snm>Gozzi</snm><fnm>N</fnm></au>
    <au><snm>Bajardi</snm><fnm>P</fnm></au>
    <au><snm>Perra</snm><fnm>N</fnm></au>
  </aug>
  <source>medRxiv</source>
  <publisher>Cold Spring Harbor Laboratory Press</publisher>
  <pubdate>2021</pubdate>
</bibl>

<bibl id="B7">
  <title><p>Non-pharmaceutical interventions during the {COVID-19} pandemic: A
  review</p></title>
  <aug>
    <au><snm>Perra</snm><fnm>N</fnm></au>
  </aug>
  <source>Physics Reports</source>
  <publisher>Elsevier</publisher>
  <pubdate>2021</pubdate>
</bibl>

<bibl id="B8">
  <title><p>The hidden geometry of complex, network-driven contagion
  phenomena</p></title>
  <aug>
    <au><snm>Brockmann</snm><fnm>D</fnm></au>
    <au><snm>Helbing</snm><fnm>D</fnm></au>
  </aug>
  <source>science</source>
  <publisher>American Association for the Advancement of Science</publisher>
  <pubdate>2013</pubdate>
  <volume>342</volume>
  <issue>6164</issue>
  <fpage>1337</fpage>
  <lpage>-1342</lpage>
</bibl>

<bibl id="B9">
  <title><p>Finding disease outbreak locations from human mobility
  data</p></title>
  <aug>
    <au><snm>Schlosser</snm><fnm>F</fnm></au>
    <au><snm>Brockmann</snm><fnm>D</fnm></au>
  </aug>
  <source>EPJ data science</source>
  <publisher>Springer Berlin Heidelberg</publisher>
  <pubdate>2021</pubdate>
  <volume>10</volume>
  <issue>1</issue>
  <fpage>52</fpage>
</bibl>

<bibl id="B10">
  <title><p>High-resolution Spatio-temporal Model for County-level {COVID-19}
  Activity in the {US}</p></title>
  <aug>
    <au><snm>Zhu</snm><fnm>S</fnm></au>
    <au><snm>Bukharin</snm><fnm>A</fnm></au>
    <au><snm>Xie</snm><fnm>L</fnm></au>
    <au><snm>Santillana</snm><fnm>M</fnm></au>
    <au><snm>Yang</snm><fnm>S</fnm></au>
    <au><snm>Xie</snm><fnm>Y</fnm></au>
  </aug>
  <source>ACM Transactions on Management Information Systems (TMIS)</source>
  <publisher>ACM New York, NY</publisher>
  <pubdate>2021</pubdate>
  <volume>12</volume>
  <issue>4</issue>
  <fpage>1</fpage>
  <lpage>-20</lpage>
</bibl>

<bibl id="B11">
  <title><p>Hawkes process modeling of {COVID-19} with mobility leading
  indicators and spatial covariates</p></title>
  <aug>
    <au><snm>Chiang</snm><fnm>WH</fnm></au>
    <au><snm>Liu</snm><fnm>X</fnm></au>
    <au><snm>Mohler</snm><fnm>G</fnm></au>
  </aug>
  <source>International journal of forecasting</source>
  <publisher>Elsevier</publisher>
  <pubdate>2021</pubdate>
</bibl>

<bibl id="B12">
  <title><p>Network Self-Exciting Point Processes To Measure Health Impacts of
  {COVID-19}</p></title>
  <aug>
    <au><snm>Giudici</snm><fnm>P</fnm></au>
    <au><snm>Pagnottoni</snm><fnm>P</fnm></au>
    <au><snm>Spelta</snm><fnm>A</fnm></au>
  </aug>
  <source>Available at SSRN 3892998</source>
  <pubdate>2021</pubdate>
</bibl>

<bibl id="B13">
  <title><p>A statistical framework for the analysis of multivariate infectious
  disease surveillance counts</p></title>
  <aug>
    <au><snm>Held</snm><fnm>L</fnm></au>
    <au><snm>Höhle</snm><fnm>M</fnm></au>
    <au><snm>Hofmann</snm><fnm>M</fnm></au>
  </aug>
  <source>Statistical Modelling</source>
  <pubdate>2005</pubdate>
  <volume>5</volume>
  <fpage>187–</fpage>
  <lpage>199</lpage>
</bibl>

<bibl id="B14">
  <title><p>Spatio-temporal analysis of epidemic phenomena using the {R}
  package surveillance</p></title>
  <aug>
    <au><snm>Meyer</snm><fnm>S</fnm></au>
    <au><snm>Held</snm><fnm>L</fnm></au>
    <au><snm>H{\"o}hle</snm><fnm>M</fnm></au>
  </aug>
  <source>arXiv preprint arXiv:1411.0416</source>
  <pubdate>2014</pubdate>
</bibl>

<bibl id="B15">
  <title><p>Inferring network topology from complex dynamics</p></title>
  <aug>
    <au><snm>Shandilya</snm><fnm>SG</fnm></au>
    <au><snm>Timme</snm><fnm>M</fnm></au>
  </aug>
  <source>New Journal of Physics</source>
  <publisher>IOP Publishing</publisher>
  <pubdate>2011</pubdate>
  <volume>13</volume>
  <issue>1</issue>
  <fpage>013004</fpage>
</bibl>

<bibl id="B16">
  <title><p>Paradoxical results in perturbation-based signaling network
  reconstruction</p></title>
  <aug>
    <au><snm>Prabakaran</snm><fnm>S</fnm></au>
    <au><snm>Gunawardena</snm><fnm>J</fnm></au>
    <au><snm>Sontag</snm><fnm>E</fnm></au>
  </aug>
  <source>Biophysical journal</source>
  <publisher>Elsevier</publisher>
  <pubdate>2014</pubdate>
  <volume>106</volume>
  <issue>12</issue>
  <fpage>2720</fpage>
  <lpage>-2728</lpage>
</bibl>

<bibl id="B17">
  <title><p>Estimating the topology of complex dynamical networks by steady
  state control: Generality and limitation</p></title>
  <aug>
    <au><snm>Yu</snm><fnm>D</fnm></au>
  </aug>
  <source>Automatica</source>
  <publisher>Elsevier</publisher>
  <pubdate>2010</pubdate>
  <volume>46</volume>
  <issue>12</issue>
  <fpage>2035</fpage>
  <lpage>-2040</lpage>
</bibl>

<bibl id="B18">
  <title><p>Inferring local dynamics and connectivity of spatially extended
  systems with long-range links based on steady-state stabilization</p></title>
  <aug>
    <au><snm>Yu</snm><fnm>D</fnm></au>
    <au><snm>Parlitz</snm><fnm>U</fnm></au>
  </aug>
  <source>Physical Review E</source>
  <publisher>APS</publisher>
  <pubdate>2010</pubdate>
  <volume>82</volume>
  <issue>2</issue>
  <fpage>026108</fpage>
</bibl>

<bibl id="B19">
  <title><p>Inferring epidemic network topology from surveillance
  data</p></title>
  <aug>
    <au><snm>Wan</snm><fnm>X</fnm></au>
    <au><snm>Liu</snm><fnm>J</fnm></au>
    <au><snm>Cheung</snm><fnm>WK</fnm></au>
    <au><snm>Tong</snm><fnm>T</fnm></au>
  </aug>
  <source>PLoS One</source>
  <publisher>Public Library of Science San Francisco, USA</publisher>
  <pubdate>2014</pubdate>
  <volume>9</volume>
  <issue>6</issue>
  <fpage>e100661</fpage>
</bibl>

<bibl id="B20">
  <title><p>Efficient network reconstruction from dynamical cascades identifies
  small-world topology of neuronal avalanches</p></title>
  <aug>
    <au><snm>Pajevic</snm><fnm>S</fnm></au>
    <au><snm>Plenz</snm><fnm>D</fnm></au>
  </aug>
  <source>PLoS computational biology</source>
  <publisher>Public Library of Science San Francisco, USA</publisher>
  <pubdate>2009</pubdate>
  <volume>5</volume>
  <issue>1</issue>
  <fpage>e1000271</fpage>
</bibl>

<bibl id="B21">
  <title><p>Network reconstruction from infection cascades</p></title>
  <aug>
    <au><snm>Braunstein</snm><fnm>A</fnm></au>
    <au><snm>Ingrosso</snm><fnm>A</fnm></au>
    <au><snm>Muntoni</snm><fnm>AP</fnm></au>
  </aug>
  <source>Journal of the Royal Society Interface</source>
  <publisher>The Royal Society</publisher>
  <pubdate>2019</pubdate>
  <volume>16</volume>
  <issue>151</issue>
  <fpage>20180844</fpage>
</bibl>

<bibl id="B22">
  <title><p>Power-law models for infectious disease spread</p></title>
  <aug>
    <au><snm>Meyer</snm><fnm>S</fnm></au>
    <au><snm>Held</snm><fnm>L</fnm></au>
  </aug>
  <source>The Annals of Applied Statistics</source>
  <publisher>Institute of Mathematical Statistics</publisher>
  <pubdate>2014</pubdate>
  <volume>8</volume>
  <issue>3</issue>
  <fpage>1612</fpage>
  <lpage>-1639</lpage>
</bibl>

<bibl id="B23">
  <title><p>The P 1 P 2/D hypothesis: on the intercity movement of
  persons</p></title>
  <aug>
    <au><snm>Zipf</snm><fnm>GK</fnm></au>
  </aug>
  <source>American sociological review</source>
  <publisher>JSTOR</publisher>
  <pubdate>1946</pubdate>
  <volume>11</volume>
  <issue>6</issue>
  <fpage>677</fpage>
  <lpage>-686</lpage>
</bibl>

<bibl id="B24">
  <title><p>A universal model for mobility and migration patterns</p></title>
  <aug>
    <au><snm>Simini</snm><fnm>F</fnm></au>
    <au><snm>Gonz{\'a}lez</snm><fnm>MC</fnm></au>
    <au><snm>Maritan</snm><fnm>A</fnm></au>
    <au><snm>Barab{\'a}si</snm><fnm>AL</fnm></au>
  </aug>
  <source>Nature</source>
  <publisher>Nature Publishing Group</publisher>
  <pubdate>2012</pubdate>
  <volume>484</volume>
  <issue>7392</issue>
  <fpage>96</fpage>
  <lpage>-100</lpage>
</bibl>

<bibl id="B25">
  <title><p>Spatial networks</p></title>
  <aug>
    <au><snm>Barth{\'e}lemy</snm><fnm>M</fnm></au>
  </aug>
  <source>Physics Reports</source>
  <publisher>Elsevier</publisher>
  <pubdate>2011</pubdate>
  <volume>499</volume>
  <issue>1-3</issue>
  <fpage>1</fpage>
  <lpage>-101</lpage>
</bibl>

<bibl id="B26">
  <title><p>Atlas du {LIBAN}</p></title>
  <aug>
    <au><snm>Verdeil</snm><fnm>E</fnm></au>
    <au><snm>Faour</snm><fnm>G</fnm></au>
    <au><snm>Velut</snm><fnm>S</fnm></au>
    <au><snm>Hamz{\'e}</snm><fnm>M</fnm></au>
    <au><snm>Mermier</snm><fnm>F</fnm></au>
  </aug>
  <publisher>Presses de l'Ifpo</publisher>
  <pubdate>2007</pubdate>
</bibl>

<bibl id="B27">
  <title><p>Secondary attack rate of {COVID-19} among contacts and risk
  factors, Tamil Nadu, March--May 2020: a retrospective cohort
  study</p></title>
  <aug>
    <au><snm>Karumanagoundar</snm><fnm>K</fnm></au>
    <au><snm>Raju</snm><fnm>M</fnm></au>
    <au><snm>Ponnaiah</snm><fnm>M</fnm></au>
    <au><snm>Kaur</snm><fnm>P</fnm></au>
    <au><snm>Rubeshkumar</snm><fnm>P</fnm></au>
    <au><snm>Sakthivel</snm><fnm>M</fnm></au>
    <au><snm>Shanmugiah</snm><fnm>P</fnm></au>
    <au><snm>Ganeshkumar</snm><fnm>P</fnm></au>
    <au><snm>Muthusamy</snm><fnm>SK</fnm></au>
    <au><snm>Sendhilkumar</snm><fnm>M</fnm></au>
    <au><cnm>others</cnm></au>
  </aug>
  <source>BMJ open</source>
  <publisher>British Medical Journal Publishing Group</publisher>
  <pubdate>2021</pubdate>
  <volume>11</volume>
  <issue>11</issue>
  <fpage>e051491</fpage>
</bibl>

<bibl id="B28">
  <title><p>Community detection in graphs</p></title>
  <aug>
    <au><snm>Fortunato</snm><fnm>S</fnm></au>
  </aug>
  <source>Physics reports</source>
  <publisher>Elsevier</publisher>
  <pubdate>2010</pubdate>
  <volume>486</volume>
  <issue>3-5</issue>
  <fpage>75</fpage>
  <lpage>-174</lpage>
</bibl>

<bibl id="B29">
  <title><p>High modularity creates scaling laws</p></title>
  <aug>
    <au><snm>Grindrod</snm><fnm>P</fnm></au>
    <au><snm>Higham</snm><fnm>DJ</fnm></au>
  </aug>
  <source>Scientific reports</source>
  <publisher>Nature Publishing Group</publisher>
  <pubdate>2018</pubdate>
  <volume>8</volume>
  <issue>1</issue>
  <fpage>1</fpage>
  <lpage>-9</lpage>
</bibl>

<bibl id="B30">
  <title><p>Statistical mechanics of complex networks</p></title>
  <aug>
    <au><snm>Albert</snm><fnm>R</fnm></au>
    <au><snm>Barab{\'a}si</snm><fnm>AL</fnm></au>
  </aug>
  <source>Reviews of modern physics</source>
  <publisher>APS</publisher>
  <pubdate>2002</pubdate>
  <volume>74</volume>
  <issue>1</issue>
  <fpage>47</fpage>
</bibl>

<bibl id="B31">
  <title><p>The structure and function of complex networks</p></title>
  <aug>
    <au><snm>Newman</snm><fnm>ME</fnm></au>
  </aug>
  <source>SIAM review</source>
  <publisher>SIAM</publisher>
  <pubdate>2003</pubdate>
  <volume>45</volume>
  <issue>2</issue>
  <fpage>167</fpage>
  <lpage>-256</lpage>
</bibl>

<bibl id="B32">
  <title><p>Epidemic processes in complex networks</p></title>
  <aug>
    <au><snm>Pastor Satorras</snm><fnm>R</fnm></au>
    <au><snm>Castellano</snm><fnm>C</fnm></au>
    <au><snm>Van Mieghem</snm><fnm>P</fnm></au>
    <au><snm>Vespignani</snm><fnm>A</fnm></au>
  </aug>
  <source>Reviews of modern physics</source>
  <publisher>APS</publisher>
  <pubdate>2015</pubdate>
  <volume>87</volume>
  <issue>3</issue>
  <fpage>925</fpage>
</bibl>

<bibl id="B33">
  <title><p>Spread of epidemic disease on networks</p></title>
  <aug>
    <au><snm>Newman</snm><fnm>ME</fnm></au>
  </aug>
  <source>Physical review E</source>
  <publisher>APS</publisher>
  <pubdate>2002</pubdate>
  <volume>66</volume>
  <issue>1</issue>
  <fpage>016128</fpage>
</bibl>

<bibl id="B34">
  <title><p>Power-law distributions in empirical data</p></title>
  <aug>
    <au><snm>Clauset</snm><fnm>A</fnm></au>
    <au><snm>Shalizi</snm><fnm>CR</fnm></au>
    <au><snm>Newman</snm><fnm>ME</fnm></au>
  </aug>
  <source>SIAM review</source>
  <publisher>SIAM</publisher>
  <pubdate>2009</pubdate>
  <volume>51</volume>
  <issue>4</issue>
  <fpage>661</fpage>
  <lpage>-703</lpage>
</bibl>

<bibl id="B35">
  <title><p>Error and attack tolerance of complex networks</p></title>
  <aug>
    <au><snm>Albert</snm><fnm>R</fnm></au>
    <au><snm>Jeong</snm><fnm>H</fnm></au>
    <au><snm>Barab{\'a}si</snm><fnm>AL</fnm></au>
  </aug>
  <source>nature</source>
  <publisher>Nature Publishing Group</publisher>
  <pubdate>2000</pubdate>
  <volume>406</volume>
  <issue>6794</issue>
  <fpage>378</fpage>
  <lpage>-382</lpage>
</bibl>

<bibl id="B36">
  <title><p>Robustness of network of networks under targeted attack</p></title>
  <aug>
    <au><snm>Dong</snm><fnm>G</fnm></au>
    <au><snm>Gao</snm><fnm>J</fnm></au>
    <au><snm>Du</snm><fnm>R</fnm></au>
    <au><snm>Tian</snm><fnm>L</fnm></au>
    <au><snm>Stanley</snm><fnm>HE</fnm></au>
    <au><snm>Havlin</snm><fnm>S</fnm></au>
  </aug>
  <source>Physical Review E</source>
  <publisher>APS</publisher>
  <pubdate>2013</pubdate>
  <volume>87</volume>
  <issue>5</issue>
  <fpage>052804</fpage>
</bibl>

<bibl id="B37">
  <title><p>Cascading failures in complex networks</p></title>
  <aug>
    <au><snm>Valdez</snm><fnm>LD</fnm></au>
    <au><snm>Shekhtman</snm><fnm>L</fnm></au>
    <au><snm>La Rocca</snm><fnm>CE</fnm></au>
    <au><snm>Zhang</snm><fnm>X</fnm></au>
    <au><snm>Buldyrev</snm><fnm>SV</fnm></au>
    <au><snm>Trunfio</snm><fnm>PA</fnm></au>
    <au><snm>Braunstein</snm><fnm>LA</fnm></au>
    <au><snm>Havlin</snm><fnm>S</fnm></au>
  </aug>
  <source>Journal of Complex Networks</source>
  <publisher>Oxford University Press</publisher>
  <pubdate>2020</pubdate>
  <volume>8</volume>
  <issue>2</issue>
  <fpage>cnaa013</fpage>
</bibl>

<bibl id="B38">
  <title><p>Catastrophic cascade of failures in interdependent
  networks</p></title>
  <aug>
    <au><snm>Buldyrev</snm><fnm>SV</fnm></au>
    <au><snm>Parshani</snm><fnm>R</fnm></au>
    <au><snm>Paul</snm><fnm>G</fnm></au>
    <au><snm>Stanley</snm><fnm>HE</fnm></au>
    <au><snm>Havlin</snm><fnm>S</fnm></au>
  </aug>
  <source>Nature</source>
  <publisher>Nature Publishing Group</publisher>
  <pubdate>2010</pubdate>
  <volume>464</volume>
  <issue>7291</issue>
  <fpage>1025</fpage>
  <lpage>-1028</lpage>
</bibl>

<bibl id="B39">
  <title><p>Structural vulnerability of the {N}orth {A}merican power
  grid</p></title>
  <aug>
    <au><snm>Albert</snm><fnm>R</fnm></au>
    <au><snm>Albert</snm><fnm>I</fnm></au>
    <au><snm>Nakarado</snm><fnm>GL</fnm></au>
  </aug>
  <source>Physical review E</source>
  <publisher>APS</publisher>
  <pubdate>2004</pubdate>
  <volume>69</volume>
  <issue>2</issue>
  <fpage>025103</fpage>
</bibl>

<bibl id="B40">
  <title><p>Understanding components of mobility during the {COVID-19}
  pandemic</p></title>
  <aug>
    <au><snm>Edsberg M{\o}llgaard</snm><fnm>P</fnm></au>
    <au><snm>Lehmann</snm><fnm>S</fnm></au>
    <au><snm>Alessandretti</snm><fnm>L</fnm></au>
  </aug>
  <source>Philosophical Transactions of the Royal Society A</source>
  <publisher>The Royal Society</publisher>
  <pubdate>2021</pubdate>
  <volume>380</volume>
  <issue>2214</issue>
  <fpage>20210118</fpage>
</bibl>

</refgrp>
} % end of \BMCxmlcomment
   % Bibliography file (usually '*.bib' )
% for author-year bibliography (bmc-mathphys or spbasic)
% a) write to bib file (bmc-mathphys only)
% @settings{label, options="nameyear"}

% or include bibliography directly:
% \begin{thebibliography}
% \bibitem{b1}
% \end{thebibliography}

%%%%%%%%%%%%%%%%%%%%%%%%%%%%%%%%%%%
%%                               %%
%% Figures                       %%
%%                               %%
%% NB: this is for captions and  %%
%% Titles. All graphics must be  %%
%% submitted separately and NOT  %%
%% included in the Tex document  %%
%%                               %%
%%%%%%%%%%%%%%%%%%%%%%%%%%%%%%%%%%%

%%
%% Do not use \listoffigures as most will included as separate files
\end{backmatter}
\end{document}